\documentclass[floatfix,twocolumn,showpacs,preprintnumbers,amsmath,amssymb,prr,superscriptaddress,longbibliography, nofootinbib]{revtex4-1}
\usepackage{color}
\usepackage[usenames,dvipsnames,svgnames,table]{xcolor}
\usepackage[colorlinks=true,linkcolor=blue,urlcolor=blue,citecolor=blue]{hyperref}
\usepackage{mathtools}
\usepackage{graphicx}
\usepackage{dcolumn}
\usepackage{array}
\usepackage{lipsum}
\usepackage{bm}
\usepackage[caption=false]{subfig}

\usepackage{amssymb}
\usepackage{multirow}
\usepackage{tabularx}
\usepackage{amsmath}
\usepackage{braket}
\usepackage{csquotes}
\graphicspath{{plots/}}
 \usepackage{lipsum}
\usepackage{mathrsfs}
\usepackage{MnSymbol}
\usepackage{extarrows}
\newcommand{\beq}{\begin{equation}}
\newcommand{\eeq}{\end{equation}}
\newcommand{\bea}{\begin{eqnarray}}
\newcommand{\eea}{\end{eqnarray}}

\renewcommand{\vec}[1]{\mathbf{#1}}

\begin{document}

%\title{Linear Response Theory with Non-local Exchange-Correlation Functionals Across Thermodynamic Regimes}

\title{Nonempirical Time-Dependent Density Functional Theory Framework for Nonlocal Exchange--Correlation Potentials}

%\title{Non-local exchange--correlation potentials in time-dependent density functional theory}

%\title{How to use non-local exchange--correlation potentials in time-dependent density functional theory}

%\title{Implications of non-local potentials in time-dependent density functional theory}

%\title{Consistent time-dependent density functional theory simulations with non-local exchange-correlation and pseudo potentials}

%\title{Non-empirical time-dependent density functional theory framework\\ for non-local potentials}

\author{Zhandos~A.~Moldabekov}
\email{z.moldabekov@hzdr.de}

\affiliation{Institute of Radiation Physics, Helmholtz-Zentrum Dresden-Rossendorf (HZDR), D-01328 Dresden, Germany}

\author{Michele Pavanello}
\affiliation{Department of Physics, Rutgers University, NJ 07102 Newark, USA}

\author{Thomas D. Gawne}
\affiliation{Center for Advanced Systems Understanding (CASUS), Helmholtz-Zentrum Dresden-Rossendorf (HZDR), D-02826 G\"orlitz, Germany}

\author{Jan Vorberger}
\affiliation{Institute of Radiation Physics, Helmholtz-Zentrum Dresden-Rossendorf (HZDR), D-01328 Dresden, Germany}

\author{Tobias Dornheim}
\email{t.dornheim@hzdr.de}
\affiliation{Institute of Radiation Physics, Helmholtz-Zentrum Dresden-Rossendorf (HZDR), D-01328 Dresden, Germany}

\affiliation{Center for Advanced Systems Understanding (CASUS), Helmholtz-Zentrum Dresden-Rossendorf (HZDR), D-02826 G\"orlitz, Germany}

%\date{\today}

\begin{abstract}
Advanced, orbital-dependent exchange--correlation (XC) functionals can significantly improve the description of electronic structural properties, but they substantially worsen spectral properties that are computed within standard linear-response time-dependent density functional theory (TDDFT) frameworks.
This is not a failure of the underlying Kohn-Sham states, but due to a formal inconsistency in the treatment of the dynamic density response when the non-locality of the XC potential is not taken into account consistently on the level of the full off-diagonal density matrix.
To avoid these complexities, we present a non-empirical additive correction $\Delta f_\textnormal{xc}(\mathbf{q},\omega)$ to the dynamic XC kernel that re-enforces the exact f-sum rule of the non-local KS Hamiltonian within TDDFT. 
Comparing our new results against a representative set of accurate experimental measurements (ambient aluminum, silicon and carbon, as well as heated and compressed aluminum) reveals a dramatic improvement in all cases without any additional computational cost. The corresponding extension to the open-source \texttt{GPAW} code is made freely available online. 
We further investigate the implications of non-local pseudopotentials and show that the non-locality has an important, physically motivated effect that is indispensable to capture the correct plasmon dispersion in lieu of full all-electron simulations. %We demonstrate that this opens up the possibility to directly assess the accuracy of any given pseudopotential against experimental measurements for arbitrary materials.
%\textcolor{red}{To be revised.}We investigate the effects of non-local exchange--correlation (XC) potentials and pseudopotentials onto the estimation of dynamic properties using linear-response time-dependent density functional theory (TDDFT). Our analysis reveals a formal inconsistency of meta-GGA and hybrid XC functionals within standard TDDFT frameworks, leading to a drastic deterioration of the computed spectra. To overcome this limitation, we introduce a non-empirical dynamic correction to the XC-kernel $f_\textnormal{xc}(\mathbf{q},\omega)$ derived from the universal f-sum rule. Comparison with experimental x-ray Thomson scattering (XRTS) measurements demonstrates the utility of our approach for three representative ambient and warm dense matter systems.
In addition to being important for the estimation of a plethora of dynamic and spectral properties, our work constitutes an important step towards a universal XC functional that can be used to estimate all kinds of observables with high accuracy.
Finally, we outline the potential utility of our framework for the development of advanced non-local XC functionals,
%constraints for the development of new non-local XC functionals, 
and suggest a new way to rigorously verify non-local pseudopotentials against experimental measurements of collective excitations. %ignatures.%dynamic response measurements.
\end{abstract}

\maketitle

% ============================================================
\section{Introduction}
% ============================================================

%\textcolor{red}{Some modification in introduction/thread/narrative to harmonically incorporate the non-local pseudo potential}

Over the last decades, density functional theory (DFT) has become established as the most successful tool for electronic-structure simulations of real materials in a great variety of contexts~\cite{Jones_RMP_2015,Jain2016,Marzari2021,Huang2023}.
Its characteristic balance between computational efficiency and accuracy is based on the formally exact mapping of the complicated quantum many-electron system of interest onto a  single-electron problem governed by the effective Kohn-Sham (KS) Hamiltonian $\hat{H}_\textnormal{KS}$~\cite{Hohenberg_Kohn_1964,Kohn_1965_A1133}.
The crucial ingredient is given by the exchange--correlation (XC) functional, which has to be approximated in practice, and which decisively determines the accuracy of a given DFT simulation~\cite{10.1039/c7cp04913g}.
Starting from relatively simple local density approximation (LDA)~\cite{Perdew1981PZ,vwn} and generalized gradient approximation (GGA)~\cite{PBE} type functionals, there has been remarkable recent progress on the development of more advanced XC functionals on higher rungs of Jacob's ladder of functional approximations~\cite{Perdew_AIP_2001}.
For example, meta-GGA and hybrid XC functionals that explicitly depend on the KS orbitals have been shown to give a vastly improved description of some important material properties such as band gaps in semiconductors and insulators~\cite{HSE03,Matsushita2011,Wing2021,Yang2023}.

\begin{figure}[t]
\centering
\includegraphics[width=0.84\linewidth]{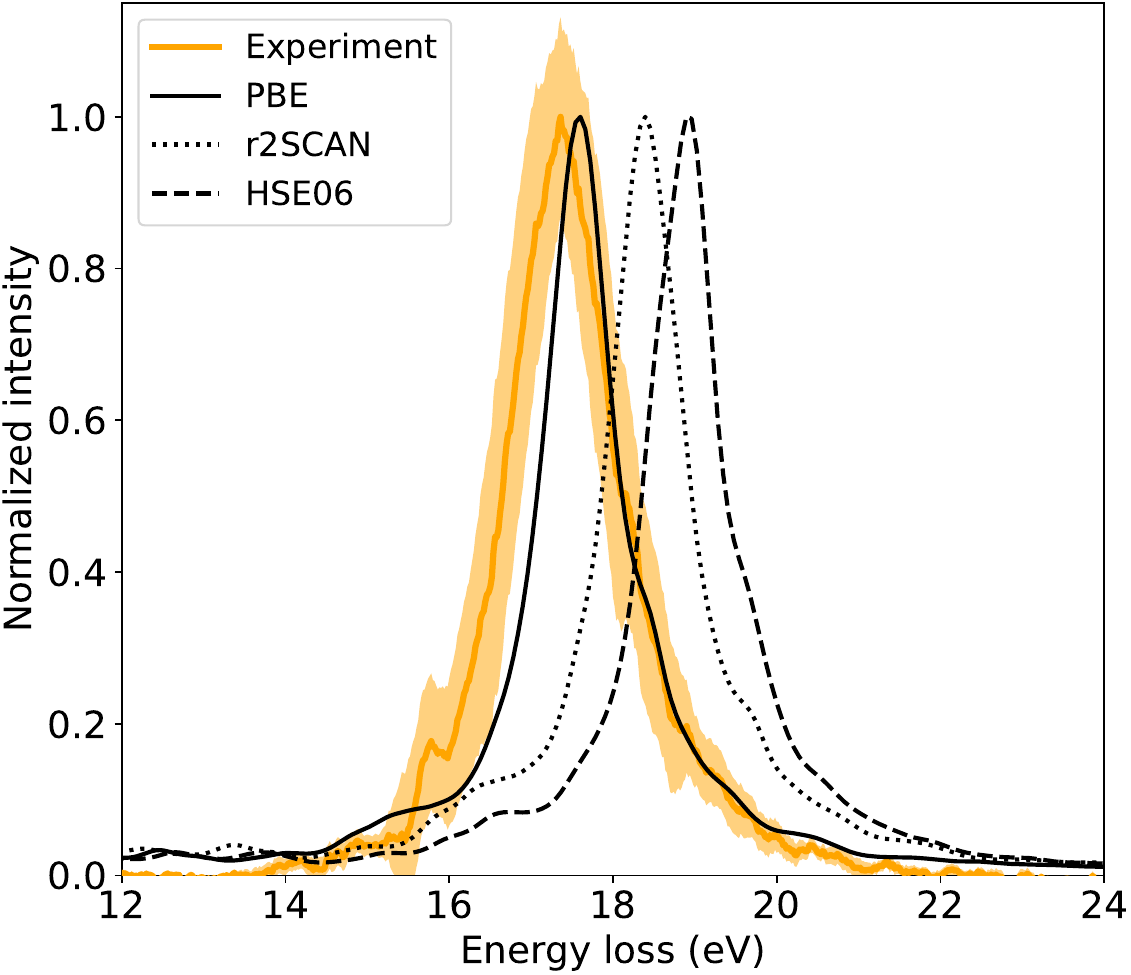}
\caption{
Plasmon of ambient aluminum with an fcc crystal structure at $q=0.93\,$\AA$^{-1}$. Orange: XRTS measurement at the European XFEL taken from Gawne \emph{et al.}~\cite{Gawne_PRB}, with the shaded area showing the corresponding statistical uncertainty of the signal. The solid, dotted and dashed black lines have been computed using TDDFT (see App.~\ref{app:Comp_det} for details) using the GGA level PBE~\cite{PBE}, meta-GGA level r2SCAN~\cite{r2SCAN} and hybrid HSE06~\cite{HSE06} XC functional. Clearly, the two non-local XC-functionals lead to a significantly poorer agreement with the experimental curve, which is being explained and corrected throughout the remainder of this work.
}
\label{fig:DSF_Al__XRTS}
\end{figure}

Recently, there has also emerged an increased interest in the properties of matter at extreme densities, temperatures and pressures. Such warm dense matter (WDM)~\cite{vorberger2025roadmapwarmdensematter} naturally occurs in a plethora of astrophysical objects such as giant planet interiors and brown dwarfs, and is also of key importance for cutting-edge technological applications such as inertial fusion energy (IFE)~\cite{Betti2016,Betti2023,Abu-Shawareb_2024} and the discovery and synthesis of novel materials~\cite{Lazicki2021,Kraus2016}.
In practice, the accurate description of WDM systems using thermal DFT methods~\cite{Mermin_DFT_1965} requires the development of corresponding thermal XC functionals, that explicitly take into account the dependence on the temperature and which must include an additional entropic XC contribution~\cite{Sjostrom_PRB_2014,ksdt,groth_prl,karasiev_importance,kushal,Moldabekov_JCTC_2024}.
Here, too, recent seminal works by Karasiev and co-workers~\cite{Mihaylov2020,Ellaboudy2025,Karasiev2026} and others~\cite{Witte_PRL,Moldabekov2023}
have highlighted the importance of advanced, non-local XC functionals.

A particularly important class of observables that can be computed using time-dependent DFT (TDDFT) methods~\cite{book_Ullrich} is given by spectral properties such as electrical and thermal conductivity and opacity, which are of key importance for the modeling of both IFE experiments and astrophysical objects~\cite{vorberger2025roadmapwarmdensematter,wdm_book}.
In addition, the dynamic density response function can be obtained using TDDFT~\cite{Schoerner_PRE_2023,Baczewski_prl_2016,Ramakrishna_PRB_2021,BespalovPRL,Moldabekov2026}. The latter is directly probed in x-ray Thomson scattering (XRTS) experiments~\cite{siegfried_review} via the electronic dynamic structure factor (DSF) $S(\mathbf{q},\omega)$ at finite momentum transfer $q=|\mathbf{q}|$. Due to this excellent point of contact between experiment and theory, XRTS has emerged as a standard method of diagnostics for extreme states of matter~\cite{dornheim2026overviewxraythomsonscattering,Tilo_Nature_2023,Dornheim_Nat_Com_2025,Kraus2026}, creating a high demand for accurate \emph{ab initio} simulations. 
Spectral measurements such as electron energy loss spectroscopy (EELS)~\cite{EELS_PRB_1989} and XRTS constitute an important way to probe the electronic structure of a given system at ambient conditions as well, allowing for more detailed comparisons between theory and experiment compared with integrated thermodynamic properties or lattice constants.

In this work, we rigorously examine the 
implications of using non-local XC functionals and non-local pseudopotentials within standard TDDFT set-ups.
For example, non-local XC functionals---commonly considered as a gold standard for the estimation of a wide range of electronic structure properties---lead to a drastically decreased quality in spectral observables compared to the simpler and computationally less expensive LDA or GGA type functionals without further corrections  \cite{Paier_PRB_2008}.
A representative example is depicted in Fig.~\ref{fig:DSF_Al__XRTS}, where we compare a high-resolution XRTS spectrum of the plasmon of ambient aluminum at $q=0.93\,$\AA$^{-1}$ obtained at the European XFEL in Germany~\cite{Gawne_PRB} (orange) with linear-response TDDFT results obtained using the GGA-level PBE~\cite{PBE} (solid black), meta-GGA level r2SCAN~\cite{r2SCAN} (dotted black) and hybrid level HSE06~\cite{HSE06} (dashed black).
Clearly, the comparably simple PBE calculation is in good agreement with the experimental spectrum, whereas r2SCAN and in particular HSE06 overestimate the true plasmon shift by more than $1.5\,$eV.
This is not an inherent deficiency in the underlying KS orbitals, but due to the inconsistent treatment of
%can be rigorously traced back to 
the non-locality of the respective XC potentials in standard TDDFT implementations.
Conversely, we show that, in lieu of computationally challenging all-electron calculations, a certain degree of non-locality is indispensable to estimate the correct plasmon excitation in the long wavelength limit. This insight opens up a new way for the experimental assessment of different non-local pseudo potentials. %%In addition, this finding points to practical limitations of time-dependent orbital-free DFT~\cite{Collins_PRL_2018, PhysRevX.11.011049, Jiang_PRB, White_PRB_2018} in its commonly used formulation employing local pseudopotentials, which provides an efficient framework for large-scale simulations of dynamic properties such as electronic stopping power~\cite{Collins_PRL_2018}.
In addition, this finding points to limitations of time-dependent orbital-free DFT~\cite{Collins_PRL_2018, PhysRevX.11.011049, Jiang_PRB, White_PRB_2018}, which commonly employs local pseudopotentials for the electron--ion interaction, when applied to the description of plasmon excitations. This has practical relevance for warm dense matter and IFE applications because orbital-free DFT provides an efficient framework for large-scale simulations of dynamic properties, including electronic stopping power~\cite{Collins_PRL_2018}.

Throughout the remainder of this work, we methodically disentangle these complexities, leading to the following results: %and derive a non-empirical dynamic correction to the XC kernel $f_\textnormal{xc}(\mathbf{q},\omega)$
%that, by enforcing the universal f-sum rule, removes the inconsistencies due to the non-locality of the utilized XC functional, leading to dramatically improved TDDFT spectra.
%In particular, we are able to report the following key results:
\begin{itemize}
    \item Non-local contributions to the effective KS Hamiltonian $\hat{H}_\textnormal{KS}$ from both the pseudopotential and/or from the XC functional lead to a modified f-sum rule with direct and important implications for the estimated spectrum and for the plasmon dispersion.
    \item Within a pseudized description of the electron--ion interaction, non-locality of the pseudopotential is essential to estimate the correct plasmon dispersion. Moreover, we show how any non-local pseudopotential gives rise to an effective electron--ion coupling energy $\Delta_\textnormal{ion}$ that is directly related to the plasmon shift and, thus, can be measured in XRTS \cite{ Laso_Garcia_2026, Gawne_PRB} or EELS~\cite{Egerton_2009} experiments.
    \item The combination of orbital-dependent non-local XC potentials (meta-GGA or hybrid level XC functionals) with standard TDDFT implementations leads to systematic inconsistencies, resulting in biased spectra and unphysical plasmon shifts.
    \item Re-enforcing the exact $f$-sum rule of the non-local KS Hamiltonian introduces a non-empirical dynamic correction $\Delta f_\textnormal{xc}(\mathbf{q},\omega)$ to the XC kernel $f_\textnormal{xc}(\mathbf{q},\omega)$ that removes this inconsistency for both static (adiabatic) and dynamic (non-adiabatic) XC kernels without additional computational cost. A corresponding implementation for the open-source \texttt{GPAW}~\cite{GPAW_2024} code is being made freely available~\cite{GitLab_codes}.
    \item We demonstrate the value of the new framework by rigorous comparisons with high-precision XRTS and EELS measurements for a number of representative examples: ambient aluminum~\cite{Gawne_PRB} and warm dense aluminum~\cite{BespalovPRL}, as well as ambient silicon~\cite{Gawne_2025} and carbon~\cite{Waidmann_EELS_Carbon}, i.e., two metallic systems and two semiconductors. In all cases, applying $\Delta f_\textnormal{xc}(\mathbf{q},\omega)$ gives drastically improved spectra and good agreement with the experimental data for the considered meta-GGA and hybrid XC functionals.
   % \item Experimental assessment of different non-local pseudopotentials, demonstrated for Al!
   % \item Way to assess plasmon directly from equilibrium DFT without need for LR-TDDFT
\end{itemize}

In this way, our work opens up a gamut of new possibilities.
First, our findings constitute an important step towards a future DFT framework in which all observables, static and dynamic, can be obtained from a single universal XC functional with comparable, high accuracy; clearly, such a functional can be expected to have a general non-local form.
This is particularly important for the description of complex WDM systems, where the application of hybrid functionals has already been shown to be important~\cite{Mihaylov2020,Ellaboudy2025,Karasiev2026,Witte_PRL,Moldabekov2023}, and where recent developments extend the applicability of TDDFT calculations to increasingly high temperatures~\cite{Moldabekov2026, BespalovPRL, White_2025, Baczewski_prl_2016}.
Conversely, our work provides additional exact constraints that are important to guide the development of advanced non-local XC functionals, with the adiabatic connection formula~\cite{pribram}
or data driven machine--learning representations~\cite{Dick2020,Kirkpatrick2021} being two promising routes. 
Furthermore, we show how plasmon spectra measured by XRTS and EELS can guide and constrain the construction of nonlocal pseudopotentials for a given material across different degrees of compression and heating. This is particularly important for WDM, where accurate measurements of other properties remain relatively sparse~\cite{vorberger2025roadmapwarmdensematter}.
Finally, we note that all of our considerations also hold for the optical limit (i.e., $q\to0$), with direct relevance for the TDDFT (and also Kubo-Greenwood) based estimation of conductivities, opacity and related properties~\cite{Shaffer_POP_2024, Sharma_POP_2026}.

The paper is organized as follows.
In Sec.~\ref{sec:theory}, we introduce the required theoretical background, including different representations of the DSF (\ref{sec:DSF}), the KS Hamiltonian for non-local potentials (\ref{sec:KS_Hamiltonian}), and the modification of the f-sum rule for non-local pseudopotentials (\ref{sec:pseudo_theory}) and non-local XC-potentials (\ref{sec:XC_theory}).
Sec.~\ref{subsec:non_imp_xc} introduces the new correction $\Delta f_\textnormal{xc}(\mathbf{q},\omega)$ to the dynamic XC-kernel, and Secs.~\ref{sec:plasmon_theory} and \ref{sec:all_el_f_sum_rule} juxtapose the description of collective plasmon excitations with non-local pseudopotentials and within the all-electron picture, respectively.
In Sec.~\ref{sec:applications}, we demonstrate the implications of our work by comparing DFT and TDDFT results against a representative set of experimental measurements, starting with a demonstration of the importance of non-locality to capture the correct plasmon shift on the example of ambient aluminum in Sec.~\ref{sec:results1}.
Sec.~\ref{sec:my_name_is_dark} is devoted to the utilization of non-local XC-functionals in TDDFT, and to highlight the substantial improvement of the computed spectra using $\Delta f_\textnormal{xc}(\mathbf{q},\omega)$ for ambient aluminum  (\ref{sec:appl_fccal}), warm dense aluminum (\ref{sec:Al_is_warm}), as well as the two prototypical semi-conductors silicon and carbon (\ref{sec:ok}).
The paper is concluded with a summary of our findings and a discussion of their implications for future works in Sec.~\ref{sec:discussion}.
Computational details and additional derivations can be found in Appendices \ref{app:Comp_det}-\ref{app:OF_TDDFT_dynamic_ei}.

\section{Theory\label{sec:theory}}

All derivations are presented in Hartree atomic units, i.e., $m_e=\hbar =k_\textnormal{B} = 1$.

\subsection{Dynamic structure factor and spectral representation\label{sec:DSF}}

Before turning to the details of TDDFT involving nonlocal potentials, it is useful to briefly recall the relations between the different dynamic response functions and the DSF directly probed in XRTS and EELS experiments. In particular, the spectral representations of the DSF and of the inverse dielectric function provide a useful framework for understanding the origin of the discrepancies discussed in Fig.~\ref{fig:DSF_Al__XRTS}.

We begin with the exact spectral representation of the DSF \cite{quantum_theory},
\begin{equation}
S(\bm q,\omega)
= \frac{1}{N_e}\sum_{\Lambda,\Lambda^{\prime}}
P_{\Lambda}\left|\langle\Lambda^{\prime}|
\hat{\rho}_{\bm q}|\Lambda\rangle
\right|^2\delta\left(\omega-\omega_{\Lambda^{\prime}\Lambda}\right),
\label{eq:DSF_Lehmann_compact}
\end{equation}
where
$\omega_{\Lambda^{\prime}\Lambda}=E_{\Lambda^{\prime}}-E_{\Lambda}$
denotes the transition frequency,
$P_{\Lambda}$ is the equilibrium probability of occupying the many-body
state $\Lambda$, and
$\left|\langle\Lambda^{\prime}|\hat{\rho}_{\bm q}|\Lambda\rangle\right|^2$
is the squared modulus of the corresponding transition matrix element of the density operator
$\hat\rho_{\bm q} =\sum_{k=1}^{N_e} e^{-i\bm q\cdot\hat{\bm r}_k}$. In the canonical ensemble, the equilibrium probability of the many-electron state $P_{\Lambda}$ is defined as $P_\Lambda=\frac{e^{-\beta \mathcal E_\Lambda}}{Z}$, with $Z=\sum_\Lambda e^{-\beta \mathcal E_\Lambda}$ and $\beta=1/T$ being the inverse temperature.

The DSF is directly related to the density-response function $\chi(\bm q,\omega)$ through the fluctuation--dissipation theorem. In the present
normalization~\cite{quantum_theory},
\begin{equation}
S(\bm q,\omega)
=
-\frac{1}{\pi n_0}
\frac{\operatorname{Im}\chi(\bm q,\omega)}
     {1-e^{-\beta\omega}},
\label{eq:FDT_finite_T}
\end{equation}
where $n_0$ is the mean electron density.

Within linear-response TDDFT, the microscopic density response of periodic systems is represented in
reciprocal space by the matrix $\chi_{\bm G\bm G'}(\bm k,\omega)$, where $\bm k$ lies in the first Brillouin zone, $\bm G$ and $\bm G'$ denote 
reciprocal-lattice vectors. Suppressing the common arguments $(\bm k,\omega)$ for compactness, the Dyson equation of linear response TDDFT reads \cite{book_Ullrich}
\begin{align}
\chi_{\bm G\bm G'}=\,&\chi_{0,\bm G\bm G'}
\nonumber\\
&+\sum_{\bm G_1,\bm G_2}
\chi_{0,\bm G\bm G_1}\,
f_{\mathrm{Hxc},\bm G_1\bm G_2}\,
\chi_{\bm G_2\bm G'}.
\label{eq:dyson}
\end{align}
Here, $\chi_0$ is the noninteracting KS density-response function and
$f_{\mathrm{Hxc}}=v_{\mathrm c}+f_{\mathrm{xc}}$ is the
Hartree--exchange-correlation kernel, with $v_{\mathrm c}$ the Coulomb
interaction and $f_{\mathrm{xc}}$ the exchange-correlation kernel.
For a physical momentum transfer $\bm q=\bm k+\bm G$, the dynamic
density response is given by the corresponding diagonal element,
$\chi(\bm q,\omega)=\chi_{\bm G\bm G}(\bm k,\omega)$, which remains coupled to off-diagonal $\bm G\ne\bm G'$ components through Eq.~(\ref{eq:dyson}), i.e., local-field effects.

The density-response function $\chi(\bm q,\omega)$  is in turn related to the diagonal inverse dielectric function according to
\begin{equation}
\epsilon^{-1}(\bm q,\omega)
=
1+v_c(q)\chi(\bm q,\omega).
\label{eq:epsinv_chi}
\end{equation}
%where $v_c(q)=4\pi/q^2$.

For convenience, we use the analytic continuation of the dynamic response
functions to complex frequencies in the intermediate derivation steps.
The corresponding Dyson equation
for ${\chi}$ retains the form of Eq.~(\ref{eq:dyson}) at complex
frequency $z$ with ${\rm Im}\,z>0$, since the causal time-domain response
relation underlying the Dyson equation can equivalently be
Fourier--Laplace transformed. 

The analytic continuation of the inverse dielectric function is analytic in the upper half
of the complex-frequency plane and can be represented as \cite{quantum_theory}
\begin{equation}
{\epsilon}^{-1}(\bm q,z)
=
1-\frac{1}{\pi}
\int_{-\infty}^{\infty}
d\omega'\,
\frac{\operatorname{Im}\epsilon^{-1}(\bm q,\omega')}
{z-\omega'}.
\label{eq:KK_full}
\end{equation}

Using the symmetry
$\operatorname{Im}\epsilon^{-1}(\bm q,-\omega)
=-\operatorname{Im}\epsilon^{-1}(\bm q,\omega)$,
Eq.~(\ref{eq:KK_full}) can be written as
\begin{equation}
{\epsilon}^{-1}(\bm q,z)
=
1-\frac{2}{\pi}
\int_{0}^{\infty}
d\omega\,
\frac{\omega\,\operatorname{Im}\epsilon^{-1}(\bm q,\omega)}
{z^2-\omega^2}.
\label{eq:KK_positive}
\end{equation}

Combining Eq.~(\ref{eq:KK_positive}) with the spectral representation
of the DSF, Eq.~(\ref{eq:DSF_Lehmann_compact}), gives the spectral
representation of the inverse dielectric function,
\begin{equation}
{\epsilon}^{-1}(\bm q,z)
=
1+\sum_{\lambda}
\frac{A_{\lambda}(\bm q)}
     {z^2-\omega_{\lambda}^2},
\label{eq:epsinv_spectral}
\end{equation}
where
$\sum_{\lambda}(\cdots)\equiv
\sum_{\Lambda,\Lambda'}^{\omega_{\Lambda'\Lambda}>0}(\cdots)$,
with $\lambda\equiv(\Lambda,\Lambda')$ denoting a composite transition
index and $\omega_{\lambda}=\omega_{\Lambda'\Lambda}$. The corresponding
spectral weights are
\begin{equation}
A_{\lambda}(\bm q)
=
2 n_0 v_c(q)\omega_{\lambda}
\left[1-\exp(-\beta\omega_{\lambda})\right]
\frac{P_{\Lambda}}{N_e}
\left|
\langle\Lambda'|
\hat{\rho}_{\bm q}
|\Lambda\rangle
\right|^2 .
\label{eq:A_lambda}
\end{equation}

For complex $z$ away from the real-axis spectral poles, the sum in
Eq.~(\ref{eq:epsinv_spectral}) is understood as an ordinary sum over
transitions. Its retarded boundary value is obtained by taking
$z=\omega+i0^+$ and using the Sokhotski--Plemelj relation ${1}/{x+i0^+}
=\mathcal P({1}/{x})-i\pi\delta(x)$, where $\mathcal P$ denotes the Cauchy principal value.

The spectral representation in Eq.~(\ref{eq:epsinv_spectral}) allows
the high-frequency expansion of
${\epsilon}^{-1}(\bm q,z)$ to be expressed in terms of the
moments of the DSF. This connection follows by expressing the spectral
denominator as a geometric sum
\begin{equation}
\frac{1}{z^2-\omega_\lambda^2}
=
\frac{1}{z^2}
\frac{1}{1-\omega_\lambda^2/z^2}
=
\frac{1}{z^2}
\sum_{n=0}^{\infty}
\left(\frac{\omega_\lambda^2}{z^2}\right)^n,
\label{eq:large_z_denominator}
\end{equation}
for $|z|>\max_\lambda |\omega_\lambda|$.

%%%For a spectrum containing a finite set of transitions, expansion (\ref{eq:large_z_denominator}) converges for $|z|>\max_\lambda |\omega_\lambda|$. For a realistic many-body spectrum, however, the excitation energies generally extend to arbitrarily large values, so that no finite $|z|$ satisfies this condition for all transitions simultaneously. In this case, expansion (\ref{eq:large_z_denominator}) is understood in the asymptotic sense for large $|z|$, provided that the corresponding frequency moments exist. The dielectric response can then be approximated by retaining the leading terms in inverse powers of $z$, with successive terms providing corrections of progressively higher order in $1/z^2$.

Substituting Eq.~(\ref{eq:large_z_denominator}) into
Eq.~(\ref{eq:epsinv_spectral}) gives the high-frequency expansion
\begin{equation}
{\epsilon}^{-1}(\bm q,z)
=
1+\frac{2\omega_{\rm p,0}^2}{q^2}
\sum_{j=1}^{\infty}
\frac{M_{2j-1}(\bm q)}{z^{2j}},
\label{eq:epsinv_moment_expansion_z}
\end{equation}
where $\omega_{\rm p,0}=(4\pi n_0)^{1/2}$
is the free-electron plasma frequency and $M_{2j-1}(\bm q)$ are the odd
frequency moments of the DSF,
\begin{equation}
\begin{aligned}
M_{2j-1}(\bm q)
&=
\int_{-\infty}^{\infty}
d\omega\,
\omega^{2j-1}S(\bm q,\omega)
\\[4pt]
&=
\frac{1}{N_e}
\sum_{\Lambda,\Lambda'}
P_{\Lambda}\,
\omega_{\Lambda'\Lambda}^{\,2j-1}
\left|
\langle\Lambda'|
\hat{\rho}_{\bm q}
|\Lambda\rangle
\right|^2,
\quad j=1,2,\ldots \,.
\end{aligned}
\label{eq:M_spectral}
\end{equation}

Using the complex-frequency form of Eq.~(\ref{eq:epsinv_chi}),
${\epsilon}^{-1}(\bm q,z)
=1+v_c(q){\chi}(\bm q,z)$,
Eq.~(\ref{eq:epsinv_moment_expansion_z}) gives
\begin{equation}
{\chi}(\bm q,z)
=
2n_0\sum_{j=1}^{\infty}
\frac{M_{2j-1}(\bm q)}{z^{2j}}.
\label{eq:chi_moment_expansion_z}
\end{equation}

For the noninteracting response, i.e., in the absence of screening and
XC effects, following the same steps used to obtain
Eq.~(\ref{eq:chi_moment_expansion_z}) gives
\begin{equation}
{\chi}_0(\bm q,z)
=
2n_0\sum_{j=1}^{\infty}
\frac{M_{2j-1}^{(0)}(\bm q)}{z^{2j}},
\label{eq:chinot_moment_expansion_z}
\end{equation}
where $M_{2j-1}^{(0)}(\bm q)$ are the moments computed using KS orbitals
and eigenenergies,
\begin{equation}
\begin{aligned}
M_{2j-1}^{(0)}(\bm q)
&=
\int_{-\infty}^{\infty}
d\omega\,
\omega^{2j-1}S_{0}(\bm q,\omega)
\\[4pt]
&=
\frac{1}{N_e}
\sum_{\Lambda,\Lambda'}
P_{\Lambda}^{\mathrm{KS}}\,
\left(\omega_{\Lambda'\Lambda}^{\mathrm{KS}}\right)^{2j-1}
\left|
\langle\Lambda'^{\mathrm{KS}}|
\hat{\rho}_{\bm q}
|\Lambda^{\mathrm{KS}}\rangle
\right|^2,
\end{aligned}
\label{eq:M_spectral_KS}
\end{equation}
where the noninteracting approximation for the DSF reads
\begin{equation}
S_{0}(\bm q,\omega)
=
-\frac{1}{\pi n_0}
\frac{\operatorname{Im}\chi_0(\bm q,\omega)}
     {1-e^{-\beta\omega}}.
\label{eq:FDTnot_finite_T}
\end{equation}

\subsection{Kohn--Sham Hamiltonian with non-local potentials \label{sec:KS_Hamiltonian}}

We consider $N_e$ electrons in the field of ions with fixed positions
$\{\bm R_I\}$, i.e., the ionic coordinates are treated as external
parameters within the Born--Oppenheimer approximation. The many-electron
Kohn--Sham Hamiltonian is
\begin{equation}
\hat H_{\mathrm{KS}}
=
\sum_{i=1}^{N_e}
\hat h_{\mathrm{KS}}^{(i)} .
\label{eq:HGKS_many}
\end{equation}
The one-electron Hamiltonian associated with electron $i$ is
\begin{equation}
\hat h_{\mathrm{KS}}^{(i)}
= \hat t_{\rm kin}^{(i)} + v_{\mathrm{loc}}^{(i)}(\hat{\bm r})
+ \hat v_{\mathrm{ion}}^{\mathrm{NL},(i)} +
\hat v_{\mathrm{xc}}^{\mathrm{NL},(i)},
\label{eq:hGKS}
\end{equation}
with $\hat t_{\rm kin}^{(i)} =\hat{\bm p}_i^{\,2}/2$, and superscript $(i)$ indicating that a one-electron operator act on an orbital $i$. For notational simplicity, the electron label $(i)$ on one-electron operators is omitted unless it is needed explicitly, in particular when
a sum over electron indices is written.

In Eq.~(\ref{eq:hGKS}), the local potential is the sum of the local part
of the electron--ion pseudopotential
$v_{\mathrm{ion}}^{\mathrm{loc}}(\bm r;\{\bm R_I\})$, the Hartree
potential $v_{\mathrm H}[n](\bm r)$, and the local part of the XC
potential $v_{\mathrm{xc}}^{\mathrm{loc}}[n](\bm r)$,
\begin{equation}
v_{\mathrm{loc}}(\bm r)
=
v_{\mathrm{ion}}^{\mathrm{loc}}(\bm r;\{\bm R_I\})
+
v_{\mathrm H}[n](\bm r)
+
v_{\mathrm{xc}}^{\mathrm{loc}}[n](\bm r).
\label{eq:vloc}
\end{equation}

In contrast to these multiplicative local potentials, both the
nonlocal part of the electron--ion pseudopotential $\hat v_{\mathrm{ion}}^{\mathrm{NL}}$ and a nonlocal XC potential $\hat v_{\mathrm{xc}}^{\mathrm{NL}}$ act on a one-electron wave function through an integral
kernel. In general, a nonlocal one-electron operator can therefore be
written as
\begin{equation}
(\hat v^{\mathrm{NL}}\psi)(\bm r)
=
\int d^3r'\,
v^{\mathrm{NL}}(\bm r,\bm r')
\psi(\bm r').
\label{eq:vxc_kernel}
\end{equation}

The Hamiltonian~(\ref{eq:HGKS_many}) defines the many-electron eigenstates
of the auxiliary noninteracting Kohn--Sham system,
\begin{equation}
\hat H_{\mathrm{KS}}\ket{\Lambda^{\mathrm{KS}}}
=
\mathcal E_{\Lambda}^{\mathrm{KS}}
\ket{\Lambda^{\mathrm{KS}}},
\end{equation}
and the corresponding noninteracting KS response $\chi_0$.

For the commutator of the KS Hamiltonian and density operator, we find
\begin{equation}
\langle \Lambda'^{\mathrm{KS}}|
[\hat H_{\mathrm{KS}},\hat\rho_{\bm q}]
|\Lambda^{\mathrm{KS}}\rangle
=
\left(
\mathcal E_{\Lambda'}^{\mathrm{KS}}
-
\mathcal E_{\Lambda}^{\mathrm{KS}}
\right)
\langle \Lambda'^{\mathrm{KS}}|
\hat\rho_{\bm q}
|\Lambda^{\mathrm{KS}}\rangle ,
\label{eq:matrix}
\end{equation}
which can be used for the calculations of the first moment of the non-interacting DSF by setting $j=1$ in Eq.~(\ref{eq:M_spectral_KS}).

We use the term pseudopotential in a broad practical sense to encompass pseudized electron--ion representations, including the PAW method~\cite{PAW_PRBE_1994}, norm-conserving pseudopotentials~\cite{NCPP_PRL_1979}, and ultrasoft pseudopotentials~\cite{Vanderbilt_PRB}, despite differences in their methodological foundations.
 
\subsection{F-sum rule for non-local pseudopotentials\label{sec:pseudo_theory}}

Substituting the high-frequency expansion of the interacting response,
Eq.~(\ref{eq:chi_moment_expansion_z}), into the left-hand side of the
Dyson equation, Eq.~(\ref{eq:dyson}), and the corresponding expansion
of the noninteracting response, Eq.~(\ref{eq:chinot_moment_expansion_z}),
into its right-hand side, and restricting to the diagonal component
$\bm G=\bm G'$, one obtains
\begin{equation}
\sum_{j=1}^{\infty}
\frac{M_{2j-1}(\bm q)}{z^{2j}}
=
\sum_{j=1}^{\infty}
\frac{M_{2j-1}^{(0)}(\bm q)}{z^{2j}}
+
\sum_{j=1}^{\infty}
\frac{\Delta M_{2j-1}(\bm q)}{z^{2j}}.
\label{eq:DSF_expansion}
\end{equation}
Here,
$\Delta M_{2j-1}(\bm q)$ denotes the difference between the
corresponding moments of the fully interacting DSF
$S(\bm q,\omega)$ and the noninteracting DSF
$S_{0}(\bm q,\omega)$.

Comparison of the coefficients of $1/z^2$  in Eq.~(\ref{eq:DSF_expansion}) therefore yields
\begin{equation}
M_1(\bm q)=M_1^{(0)}(\bm q)+\Delta M_1(\bm q).
\label{eq:moment_identity2}
\end{equation}

Next, we derive general analytical expressions for the two contributions to the first moment of the DSF, $M_1^{(0)}(\bm q)$ and $\Delta M_1(\bm q)$. As shown below, the standard TDDFT formulation
combined with nonlocal XC potentials leads to an inconsistency in the first moment, which results in an erroneous blue shift of the plasmon
frequency, as illustrated in Fig.~\ref{fig:DSF_Al__XRTS}.

\subsubsection{The first moment
\texorpdfstring{$M_1^{(0)}(\bm q)$}{M1(0)(q)}
of the noninteracting DSF}

For an isotropic equilibrium state that satisfies the symmetry
$S(\bm q,\omega)=S(-\bm q,\omega)$, using Eq.~(\ref{eq:M_spectral_KS}), the first moment $M_1^{(0)}(\bm q)$ can be written in terms of a
double commutator as
\begin{equation}
M_{1}^{(0)}(\bm q) =\frac{1}{2N_e} \sum_{\Lambda}
P_\Lambda \left\langle \Lambda
\left| [\hat\rho_{-\bm q}, [\hat H_{\mathrm{KS}},\hat\rho_{\bm q}]]
\right| \Lambda \right\rangle .
\label{eq:M1_double}
\end{equation}

For the Hamiltonian in Eq.~(\ref{eq:HGKS_many}), the first moment $M_1^{(0)}(\bm q)$ can
therefore be decomposed as
\begin{align}
M_{1}^{(0)}(\bm q) =\frac{q^2}{2}+\Delta_{\mathrm{ion}}(\bm q)+\Delta_{\mathrm{xc}}(\bm q),
\label{eq:double_decomposition}
\end{align}
where  the term ${q^2}/{2}$ on the right hand side of Eq.~(\ref{eq:double_decomposition}) is the standard result following from the kinetic energy operator \cite{quantum_theory}, in addition to which, we introduced the contributions due to the nonlocal part of the electron--ion pseudopotential $\hat v_{\mathrm{ion}}^{\mathrm{NL}}$,
\begin{equation}
\Delta_{\mathrm{ion}}(\bm q)
=
\frac{1}{2N_e}\sum_{i=1}^{N_e}\sum_{\Lambda}P_\Lambda
\left\langle \Lambda \left|
[\hat\rho_{-\bm q},[\hat v_{\mathrm{ion}}^{\mathrm{NL},(i)},\hat\rho_{\bm q}]]
\right| \Lambda \right\rangle ,
\label{eq:Delta_ion}
\end{equation}
 and a nonlocal XC potential $\hat v_{\mathrm{xc}}^{\mathrm{NL}}$,
\begin{equation}
\Delta_{\mathrm{xc}}(\bm q)
=
\frac{1}{2N_e}\sum_{i=1}^{N_e}\sum_{\Lambda}P_\Lambda
\left\langle \Lambda \left|
[\hat\rho_{-\bm q},[\hat v_{\mathrm{xc}}^{\mathrm{NL},(i)},\hat\rho_{\bm q}]]
\right| \Lambda \right\rangle .
\label{eq:Delta_xc}
\end{equation}

Eqs.~(\ref{eq:Delta_ion}) and (\ref{eq:Delta_xc}) can be used to calculate $\Delta_{\mathrm{ion}}(\bm q)$ and $\Delta_{\mathrm{xc}}(\bm q)$, respectively, directly from the equilibrium KSDFT wavefunctions, without performing TDDFT calculations. We provide an open-source code that uses the \texttt{GPAW} KS wavefunctions to compute $\Delta_{\mathrm{ion}}(\bm q)$ within the PAW method and $\Delta_{\mathrm{xc}}(\bm q)$ for range-separated hybrid functionals of the HSE type, and the meta-GGA functionals r2SCAN and SCAN in an online repository~\cite{GitLab_codes}.

\subsubsection{
The first moment
\texorpdfstring{$M_1(\bm q)$}{M1(0)(q)}
with \texorpdfstring{$\hat v_{\mathrm{xc}}^{\mathrm{NL}}=0$}{v_xc^NL=0}}

Next we consider the TDDFT with $\hat v_{\mathrm{xc}}^{\mathrm{NL}}=0$, i.e., with local and semi-local XC potentials. Examples of such XC potentials are the LDA and GGA. In this case, to learn about $\Delta M_1(\bm q)$, we use the expanded form of the Dyson equation, Eq.~(\ref{eq:dyson}), for $\vec G=\vec G^{\prime}$,
\begin{align}
\chi_{\bm G\bm G}
={}&
\chi_{0,\bm G\bm G}
+
\left(
\chi_0\circ f_{\mathrm{Hxc}}\circ\chi_0
\right)_{\bm G\bm G}
\nonumber\\
&+
\left(
\chi_0\circ f_{\mathrm{Hxc}}\circ\chi_0
\circ f_{\mathrm{Hxc}}\circ\chi_0
\right)_{\bm G\bm G}
+\cdots ,
\label{eq:dyson_expanded_std}
\end{align}
where $\circ$ denotes the contraction over intermediate reciprocal-lattice
indices, and the common arguments $(\bm k,z)$ of the response
functions and kernel are suppressed for notational simplicity.

To make further progress, we first recall the high-frequency behavior
of the noninteracting response,
$\chi_0\sim z^{-2}+\mathcal O(z^{-4})$.
We then consider XC kernels that remain finite in this limit.
%%%%, including static kernels and dynamic kernels constructed to satisfy the all-electron $f$-sum rule, the latter being discussed in Sec.~\ref{sec:all_el_f_sum_rule}.

For such kernels, the first kernel-dependent term in the iterative
Dyson expansion, Eq.~(\ref{eq:dyson_expanded_std}), behaves as
$\chi_0 f_{\mathrm{Hxc}}\chi_0\sim z ^{-4}$.
Consequently, the kernel does not modify the leading
$z^{-2}$ coefficient of the density response. Therefore,
for $\chi(\bm q,z)=\chi_{\bm G\bm G}(\bm k,z)$,
\begin{equation}
\frac{2n_0M_1(\bm q)}{z^2}
+\mathcal O(z^{-4})
=
\frac{2n_0M_1^{(0)}(\bm q)}{z^2}
+\mathcal O(z^{-4}),
\label{eq:moment_identity_inter}
\end{equation}
with
\begin{equation}
    \left.\Delta_{\mathrm{xc}}(\bm q)\right|_{\hat v_{\mathrm{xc}}^{\mathrm{NL}}=0}
=0, \quad  \left.\Delta M_1(\bm q)\right|_{\hat v_{\mathrm{xc}}^{\mathrm{NL}}=0}
=0.
\end{equation}

Thus, when $\hat v_{\mathrm{xc}}^{\mathrm{NL}}=0$, the first moment of the TDDFT
DSF is fully determined by the corresponding noninteracting KS density
response function:
\begin{equation}
\left.M_1(\bm q)\right|_{\hat v_{\mathrm{xc}}^{\mathrm{NL}}=0}
=M_1^{(0)}(\bm q).
\label{eq:moment1}
\end{equation}

The more involved case $\hat v_{\mathrm{xc}}^{\mathrm{NL}}\neq 0$,
corresponding to KSDFT with a nonlocal XC potential, is treated separately in Sec.~\ref{s:f_sum_nl_xc}, as the standard
density-response formulation of TDDFT must be extended to the linear
density-matrix response formalism~\cite{Furche_jcp_2001, Furche_JCTC_2023}.

\subsection{F-sum rule for non-local XC potentials\label{sec:XC_theory}}
\label{s:f_sum_nl_xc}
Next, we consider the implications of employing nonlocal XC functionals in TDDFT. In particular, we consider $\Delta M_1(\bm q)$ in the case $\hat v_{\mathrm{xc}}^{\mathrm{NL}}\neq 0$. A key point is that knowledge of the density perturbation $\delta n(\bm r)$ alone is generally insufficient to determine the variation of a nonlocal XC potential. Unlike LDA and GGA type approximations, nonlocal XC functionals depend explicitly on the KS orbitals, or equivalently on the one-particle density matrix. Their response therefore involves variations of these orbital degrees of freedom and cannot, in general, be written solely in terms of the induced density. 

%%By contrast, no analogous additional response variable arises from the nonlocal pseudopotentials considered above. The ions and the pseudopotential operators are held fixed during the electronic response, so that $\delta \hat V_{\mathrm{ion}}^{\mathrm{NL}}=0$. The nonlocal pseudopotential therefore does not generate an additional contribution to the induced KS potential. Its effect is instead fully contained in the ground-state KS orbitals and eigenvalues.

Therefore, to derive the correct result for the first moment of the DSF in the case when a non-local XC potential is used, one has to consider the more involved linear response of the full off-diagonal one-particle density-matrix.

\subsubsection{From one-particle density-matrix response to the DSF}

The one-particle density-matrix response function $L$ relates an
external one-particle perturbation $\delta v_{\mathrm{ext}}$ to the
induced change of the one-particle density matrix $\delta\gamma$. In the case of a local external potential $\delta v_{\mathrm{ext}}(\bm x,z)$ that only depends on one spatial coordinate $\bm x$, they are related by
\begin{align}
\delta\gamma(\bm r,\bm r';z)
={}&
\int d\bm x\,
L(\bm r,\bm r';\bm x,\bm x;z)
\,\delta v_{\mathrm{ext}}(\bm x,z),
\label{eq:L_definition}
\end{align}
from which the density perturbation follows by considering the diagonal part,
\begin{align}
\delta n(\bm r,z)
={}&
\int d\bm x\,
L(\bm r,\bm r;\bm x,\bm x;z)
\,\delta v_{\mathrm{ext}}(\bm x,z).
\end{align}

Similarly to the Dyson equation of linear-response TDDFT, one can write
the Dyson-type equation for the one-particle density-matrix response,
\begin{equation}
L= L_0+ L_0\circ{\mathcal K}\circ L,
\label{eq:L_Dyson}
\end{equation}
where $\circ$ denotes integration over intermediate coordinate pairs,
$L_0$ is the noninteracting density-matrix response
function (the density-matrix-response analogue of the noninteracting
density-response function ${\chi}_0$).
A detailed derivation of Eq.~(\ref{eq:L_Dyson}) is given in
App.~\ref{app:density_matrix_dyson}.

The kernel ${\mathcal K}$ in Eq.~(\ref{eq:L_Dyson}) describes
the variation of the induced one-particle potential with respect to the
density matrix and contains the Hartree contribution and the local and
nonlocal contributions from the XC potential,
\begin{equation}
{\mathcal K}
=
{\mathcal K}_{\mathrm H}
+
{\mathcal K}_{\mathrm{xc}}^{\mathrm{loc}}
+
{\mathcal K}_{\mathrm{xc}}^{\mathrm{NL}}.
\label{eq:K_decomposition}
\end{equation}
The Hartree and local XC contributions depend only on the diagonal
density variation,
$\delta n(\bm r,z)=\delta\gamma(\bm r,\bm r;z)$,
whereas ${\mathcal K}_{\mathrm{xc}}^{\mathrm{NL}}$ acts on the
full two-coordinate density-matrix variation
$\delta\gamma(\bm r,\bm r';z)$. For fixed ions,
$\delta\hat v_{\mathrm{ion}}^{\mathrm{loc}}
=\delta\hat v_{\mathrm{ion}}^{\mathrm{NL}}=0$,
and therefore the fixed ionic pseudopotential does not contribute to
${\mathcal K}$.

The density-response functions of linear-response TDDFT are contained
in the corresponding density-matrix response functions and are obtained
by taking their double-diagonal projection,
\begin{equation}
{\chi}(\bm r,\bm x;z)
=
L(\bm r,\bm r;\bm x,\bm x;z),
\end{equation}
and analogously
${\chi}_0(\bm r,\bm x;z)
=
L_0(\bm r,\bm r;\bm x,\bm x;z)$.
In reciprocal space, these functions are represented by
${\chi}_{\bm G\bm G'}(\bm k,z)$ and
${\chi}_{0,\bm G\bm G'}(\bm k,z)$, respectively.
Therefore, for the momentum transfer $\bm q=\bm k+\bm G$, the diagonal
reciprocal-space density response can be obtained directly from the
four-point density-matrix response as
\begin{equation}
{\chi}(\bm q,z)
=
{\chi}_{\bm G\bm G}(\bm k,z)
\equiv
\mathcal P_{\bm q}[L](z),
\label{eq:chi_projection}
\end{equation}
and analogously
${\chi}_0(\bm q,z)
=
\mathcal P_{\bm q}[L_0](z)$.
Here, $\mathcal P_{\bm q}$ denotes the combined projection onto the
diagonal coordinate pairs and the subsequent Fourier transformation,
\begin{equation}
\mathcal P_{\bm q}[L](z)
\equiv
\frac{1}{\Omega}
\int d\bm r\,d\bm x\,
e^{-i\bm q\cdot\bm r}
L(\bm r,\bm r;\bm x,\bm x;z)
e^{i\bm q\cdot\bm x},
\label{eq:Pq_definition}
\end{equation}
where $\Omega$ is the normalization volume.

The physical retarded density response is obtained as the boundary value
\begin{equation}
\chi(\bm q,\omega)
=
\lim_{\eta\to0^+}
{\chi}(\bm q,\omega+i\eta)
=
\lim_{\eta\to0^+}
\mathcal P_{\bm q}[L](\omega+i\eta).
\end{equation}
Together with the fluctuation-dissipation theorem in
Eq.~(\ref{eq:FDT_finite_T}), this establishes the relation between the
one-particle density-matrix response and the DSF.

\subsubsection{Nonlocal-XC correction to the first moment of the DSF}

We now consider the contribution of a nonlocal XC potential to the first moment of the DSF using  the Dyson-type equation for the one-particle density-matrix response $L$, Eq.~ (\ref{eq:L_Dyson}), and connection (\ref{eq:chi_projection}). Since the first moment is determined by the coefficient of
$1/z^2$ in the high-frequency expansion of
$\chi(\bm q,z)$, we only need to identify which terms in the
Dyson series for $L$ can contribute at this order.

Iterating Eq.~(\ref{eq:L_Dyson}) gives
\begin{equation}
L=L_0+L_0\circ\mathcal K\circ L_0
+L_0\circ\mathcal K\circ L_0\circ\mathcal K\circ L_0+\cdots .
\label{eq:L_expansion}
\end{equation}
After applying projection $\mathcal P_{\bm q}$ (\ref{eq:Pq_definition}) to expansion (\ref{eq:L_expansion}),
\begin{align}
\chi(\bm q,z)
={}&
\mathcal P_{\bm q}[L_0]
+\mathcal P_{\bm q}[L_0\circ\mathcal K\circ L_0]
\nonumber\\
&+\mathcal P_{\bm q}[L_0\circ\mathcal K\circ L_0
\circ\mathcal K\circ L_0]
+\cdots .
\label{eq:chi_L_expansion}
\end{align}
The first term in Eq.~(\ref{eq:chi_L_expansion}) is simply the noninteracting density response,
\begin{equation}
\mathcal P_{\bm q}[L_0]
=\chi_0(\bm q,z)
=\frac{2n_0M_1^{(0)}(\bm q)}{z^2}
+\mathcal O(z^{-4}),
\label{eq:PL0_hf}
\end{equation}
where $M_1^{(0)}(\bm q)$ is given by
Eq.~(\ref{eq:double_decomposition}).

The essential difference between the high-frequency behavior of the
noninteracting density response function $\chi_0$ in
Eq.~(\ref{eq:chi_moment_expansion_z}) and that of the unprojected
density-matrix response $L_0$ is that the expansion of $L_0$ starts
one order earlier, with a term proportional to $z^{-1}$:
\begin{equation}
L_0(z)=\frac{L_0^{(1)}}{z}
+\frac{L_0^{(2)}}{z^2}
+\mathcal O(z^{-3}).
\label{eq:L0hf}
\end{equation}
The derivation of Eq.~(\ref{eq:L0hf}), together with explicit expressions
for $L_0^{(1)}$ and $L_0^{(2)}$, is given in
App.~\ref{app:L0_high_frequency}.

The $1/z$ term does not appear in
$\chi_0=\mathcal P_{\bm q}[L_0]$, since
$\mathcal P_{\bm q}[L_0^{(1)}]=0$ (see App.~\ref{app:L0_high_frequency}). Nevertheless, this term contributes
to the first kernel-dependent correction $L_0\circ\mathcal K\circ L_0$ in
Eq.~(\ref{eq:chi_L_expansion}), because the kernel is inserted between
two unprojected density-matrix response functions. Consequently,
$L_0^{(1)}/z$ from each side of the kernel generates a contribution
of order $z^{-2}$ before the final projection onto the density
response is performed.

In the high-frequency regime where the inverse-frequency expansion is
applicable, we assume that the kernel remains bounded and admits a
regular expansion with a frequency-independent leading term,
\begin{equation}
\mathcal K(z)=
\mathcal K^{(0)}+\mathcal O(z^{-1}).
\label{eq:K_hf}
\end{equation}
This behavior is consistent with conventional interaction kernels, obtained after projection onto the density response, whose frequency dependence arises from finite-memory effects~\cite{Hong_PRL, Hong_PRA, Vorberger_PRL, hansen2013theory}.

The first kernel insertion then behaves as
\begin{equation}
L_0\circ\mathcal K\circ L_0
=\frac{L_0^{(1)}\circ\mathcal K^{(0)}\circ L_0^{(1)}}{z^2}
+\mathcal O(z^{-3}).
\label{eq:L0KL0_hf}
\end{equation}
It can therefore contribute directly to the same $1/z^2$
coefficient that determines the first moment. By contrast, the next
iteration satisfies
\begin{equation}
L_0\circ\mathcal K\circ L_0
\circ\mathcal K\circ L_0
=
\mathcal O(z^{-3}),
\label{eq:L0KL0KL0_hf}
\end{equation}
and every additional kernel insertion introduces at least one further
inverse power of $z$. Since the projection $\mathcal P_{\bm q}$
is frequency independent, it 
%can eliminate a given asymptotic coefficient but 
cannot convert a term of order $z^{-3}$ or
higher into one of order $z^{-2}$.

Consequently, the coefficient of the leading $1/z^2$ term in the
full density response receives two contributions: the direct
noninteracting contribution from $L_0$ and the contribution generated
by the first kernel insertion. Thus,
\begin{align}
\chi(\bm q,z)
={}&
\frac{2n_0M_1^{(0)}(\bm q)}{z^2}
+
\frac{1}{z^2}
\mathcal P_{\bm q}
\left[
L_0^{(1)}
\circ\mathcal K^{(0)}
\circ L_0^{(1)}
\right]
+
\mathcal O(z^{-3}).
\label{eq:chi_hf_density_matrix}
\end{align}
On the other hand, the general high-frequency expansion of the density
response, Eq.~(\ref{eq:chi_moment_expansion_z}), leads to
\begin{equation}
\chi(\bm q,\omega)
=\frac{2n_0M_1(\bm q)}{z^2}
+\mathcal O(z^{-4}).
\label{eq:chi_M1_repeat}
\end{equation}
Equating the coefficients of $1/z^2$ from Eq.~(\ref{eq:chi_hf_density_matrix}) and Eq.~(\ref{eq:chi_M1_repeat})  gives
\begin{equation}
M_1(\bm q)
=M_1^{(0)}(\bm q)
+\Delta M_1(\bm q),
\label{eq:M1_from_L}
\end{equation}
where the contribution generated by the density-matrix kernel is
\begin{equation}
\Delta M_1(\bm q)
=
\frac{1}{2n_0}
\mathcal P_{\bm q}
\left[
L_0^{(1)}
\circ\mathcal K^{(0)}
\circ L_0^{(1)}
\right].
\label{eq:DeltaM1_kernel}
\end{equation}
Finally, using the decomposition of the noninteracting first moment
from Eq.~(\ref{eq:double_decomposition}), $M_1^{(0)}(\bm q)
={q^2}/{2}+\Delta_{\mathrm{ion}}(\bm q)
+\Delta_{\mathrm{xc}}(\bm q)$,
we obtain
\begin{equation}
M_1(\bm q)
=\frac{q^2}{2}
+\Delta_{\mathrm{ion}}(\bm q)
+\Delta_{\mathrm{xc}}(\bm q)
+\Delta M_1(\bm q).
\label{eq:M1_L0}
\end{equation}

Eq.~(\ref{eq:DeltaM1_kernel}) is not a first-order approximation in the kernel, but an exact expression for $\Delta M_1(\bm q)$. All terms containing two or more kernel insertions contribute only beyond order $1/z^2$ and therefore cannot modify
$M_1(\bm q)$.

At this stage, Eq.~(\ref{eq:DeltaM1_kernel}) appears to imply that $\Delta M_1(\bm q)$ must be worked out from the detailed form of the
nonlocal XC potential and its density-matrix kernel for each functional separately. A less obvious possibility emerges, however, when one asks
whether a fundamental symmetry constrains this contribution independently of those details. A key observation of the present work is that local gauge covariance provides precisely such a constraint, allowing $\Delta M_1(\bm q)$ to be determined without specifying the detailed form of the nonlocal XC functional or its kernel.
Specifically, if the nonlocal XC potential and its associated density-matrix kernel transform consistently under a local gauge transformation, local gauge covariance requires
\begin{equation}
\Delta M_1(\bm q)
=-\Delta_{\mathrm{xc}}(\bm q),
\label{eq:xc_cancellation_condition}
\end{equation}
where $\Delta_{\mathrm{xc}}(\bm q)$ is defined in
Eq.~(\ref{eq:Delta_xc}). A proof of this identity is given in App.~\ref{app:xc_gauge_cancellation}.

Eq.~(\ref{eq:xc_cancellation_condition}) has a direct and important
consequence for the physical first moment. Although the nonlocal XC
potential modifies the first moment of the noninteracting KS DSF, $M_1^{(0)}(\bm q)$,
through $\Delta_{\mathrm{xc}}(\bm q)$ (see Eq.~(\ref{eq:double_decomposition})), a gauge-consistent density-matrix response generates precisely the compensating contribution $-\Delta_{\mathrm{xc}}(\bm q)$. The nonlocal XC
contribution therefore cancels identically from the first moment of the
interacting DSF, yielding
\begin{align}
M_1(\bm q)
=\frac{q^2}{2}
+\Delta_{\mathrm{ion}}(\bm q).
\label{eq:M1_exact}
\end{align}

For the Hartree--Fock exact-exchange kernel within KSDFT, the vanishing
net contribution of the nonlocal XC potential to the $f$-sum rule was
previously demonstrated in Ref.~\cite{Hellgren_PRB2008} through an
explicit term-by-term analysis within full many-body perturbation theory.
This finding is consistent with the criterion derived here, since
Hartree--Fock exact exchange satisfies local gauge covariance~\cite{Epstein_JCP_1965}. We emphasize that the present analysis extends this conclusion beyond that specific case. 
Since exact Hartree--Fock exchange with either bare or screened Coulomb interactions is locally gauge covariant, this property is retained by hybrid functionals such as HSE06~\cite{HSE06_Krukau} and PBE0~\cite{PBE0}, which incorporate a fraction of Hartree--Fock exchange.
In contrast, some meta-GGA functionals that depend on the kinetic-energy density, $\tau(\bm r,t)=\frac{1}{2}\sum_i f_i\left|\nabla\varphi_i(\bm r,t)\right|^2$, such as  SCAN \cite{SCAN}, have been shown to violate this condition \cite{Richter_JCP_2023, Furche_JCP_2022}.

For an all-electron Hamiltonian with the local Coulomb electron--ion
interaction (hence $\Delta_{\mathrm{ion}}(\bm q)=0$), Eq.~(\ref{eq:M1_exact}) reduces to the familiar form of
the $f$-sum rule,
\begin{equation}
M_1^{\rm AE}(\bm q)=\frac{q^2}{2},
\end{equation}
where the superscript ``AE'' denotes the all-electron result.

It is known that the use of nonlocal pseudopotentials generally leads to a deviation of the first moment from $q^2/2$~\cite{Levine_prl_1989, Yabana_PRB_2000, Alippi_PRB_1997}, and this deviation is often described as a violation of the
$f$-sum rule~\cite{TIMROV2015460,turboTDDFT}. Here, however, we arrive at a fundamentally different conclusion. When nonlocal pseudopotentials are used, the deviation of the first moment from $q^2/2$ through $\Delta_{\mathrm{ion}}(\bm q)$ is not a
violation of the $f$-sum rule. Instead, it is the physically required modification of the first-moment relation generated by the nonlocal
electron--ion interaction. This distinction is not merely semantic; it has a direct physical
consequence. As shown in Sec.~\ref{sec:plasmon_theory},
$\Delta_{\mathrm{ion}}(\bm q)$ determines a contribution to the
plasmon-frequency shift relative to the free-electron-gas value, which
can be probed experimentally, e.g., by XRTS as demonstrated in Sec.~\ref{sec:results1}.

When the KS states are generated from a Hamiltonian containing a nonlocal XC potential,
$\hat v_{\mathrm{xc}}^{\mathrm{NL}}\neq 0$, the conventional density-response formulation of TDDFT does not, in general, recover the first-moment relation in Eq.~(\ref{eq:M1_exact}) when used with static XC kernels or with dynamic kernels constrained by the all-electron $f$-sum rule. The reason is that the standard TDDFT, which is restricted to the diagonal density response, does not cancel the nonlocal XC contribution and gives $\Delta M_1(\bm q)\equiv0$ in Eq.~(\ref{eq:M1_L0}). A consistent interacting response must therefore generate the
corresponding contribution $\Delta M_1(\bm q)=-\Delta_{\mathrm{xc}}(\bm q)$ from the same XC
functional, so that the two terms cancel in the physical first moment. This compensating contribution arises naturally in the one-particle
density-matrix response formalism considered above.
Such calculations are, however, considerably more computationally demanding than the conventional density-response formulation. This becomes particularly restrictive for warm dense matter and dense plasmas, where thermal excitation leads to a large number of partially occupied states. To our knowledge, an explicit density-matrix response
treatment of this type has not yet been applied to such systems.

We therefore derive below a dynamic correction to the conventional XC kernel that accounts for the missing $-\Delta_{\mathrm{xc}}(\bm q)$ contribution within the standard density-response formulation of TDDFT. The resulting correction restores the exact first-moment relation in Eq.~(\ref{eq:M1_exact}) without requiring an explicit density-matrix response calculation.

\subsection{Non-empirical correction for dynamic XC kernel}\label{subsec:non_imp_xc}

For the XRTS and EELS applications considered here, we first formulate
the correction directly in terms of the density response. This allows
the first-moment constraint in Eq.~(\ref{eq:M1_exact}) to be enforced
within conventional linear-response TDDFT without modifying the
underlying Dyson-equation solver. The corresponding formulation for the
full response matrix $\chi_{\bm G\bm G'}(\bm k,\omega)$ is given in
App.~\ref{app:dynamic_kernel_matrix}.

To determine the frequency dependence required by the first-moment
constraint, we first consider the analytically continued response in the
asymptotic region of the upper complex-frequency plane with ${\rm Im}\,z>0$ while remaining separated
from the real-frequency axis. 

Making use of the general structure of the relation between an
XC kernel and the inverse density response, we write
\begin{equation}
\chi^{-1}(\bm q,z)
=
\chi_{\rm in}^{-1}(\bm q,z)
-
\Delta f_{\mathrm{xc}}(\bm q,z),
\label{eq:chi_corr}
\end{equation}
where the subscript ``in'' denotes the response corresponding to linear-response TDDFT prior to the inclusion of the dynamic correction.

Using $M_1^{(0)}(\vec q)$ as the first DSF moment associated with $\chi_{\rm in}(\bm q,z)$, from Eq.~(\ref{eq:chi_moment_expansion_z}), the leading asymptotic term of the input response is
\begin{equation}
\chi_{\rm in}(\bm q,z)
=
\frac{2n_0M_1^{(0)}(\bm q)}{z^2}
+\mathcal O(z^{-4}).
\label{eq:chi_in_asymptotic}
\end{equation}

Since Eq.~(\ref{eq:chi_in_asymptotic}) scales as $1/z^2$, the minimal
analytic correction in Eq.~(\ref{eq:chi_corr}) capable of modifying the
coefficient of the leading $\chi(\bm q,z)\sim 1/z^2$ term is
\begin{equation}
\Delta f_{\mathrm{xc}}(\bm q,z)=C(\bm q)z^2 .
\label{eq:dynamic_xc_ansatz_z}
\end{equation}

Rearranging Eq.~(\ref{eq:chi_corr}) and substituting
Eqs.~(\ref{eq:chi_in_asymptotic}) and
(\ref{eq:dynamic_xc_ansatz_z}) gives
\begin{equation}
\chi(\bm q,z)
=
\frac{2n_0M_1^{(0)}(\bm q)}
{1-2n_0C(\bm q)M_1^{(0)}(\bm q)}
\frac{1}{z^2}
+\mathcal O(z^{-4}).
\label{eq:chi_corr_asymptotic}
\end{equation}
Matching the leading coefficient in Eq.~(\ref{eq:chi_corr_asymptotic})
to that in Eq.~(\ref{eq:chi_moment_expansion_z}) yields
\begin{equation}
C(\bm q)
=
\frac{1}{2n_0}
\left[
\frac{1}{M_1^{(0)}(\bm q)}
-
\frac{1}{M_1(\bm q)}
\right].
\label{eq:C_xc}
\end{equation}

Using Eqs.~(\ref{eq:M1_exact}) and
(\ref{eq:double_decomposition}), Eq.~(\ref{eq:C_xc}) becomes
\begin{equation}
C(\bm q)
=
-\frac{\Delta_{\mathrm{xc}}(\bm q)}
{2n_0M_1^{(0)}(\bm q)
\left[{q^2}/{2}+\Delta_{\mathrm{ion}}(\bm q)\right]}.
\label{eq:C_xc_explicit}
\end{equation}

The asymptotic analysis above is used to determine the coefficient $C(\bm q)$  in
Eq.~(\ref{eq:dynamic_xc_ansatz_z}). Having fixed
$C(\bm q)$, we return to the physical retarded response by taking the
boundary value of the correction,
\begin{equation}
\Delta f_{\mathrm{xc}}(\bm q,\omega)
=
\lim_{\eta\to0^+}
\Delta{f}_{\mathrm{xc}}
(\bm q,\omega+i\eta)
=
C(\bm q)\omega^2.
\end{equation}
Thus, the dynamic correction for the real-frequency TDDFT calculations is
\begin{align}
\Delta f_{\mathrm{xc}}(\bm q,\omega)
={}&
-\frac{\Delta_{\mathrm{xc}}(\bm q)}
{2n_0M_1^{(0)}(\bm q)
\left[{q^2}/{2}+\Delta_{\mathrm{ion}}(\bm q)\right]}
\,\omega^2,
\label{eq:dynamic_xc_correction}
\end{align}
where
$M_1^{(0)}(\bm q)
={q^2}/{2}+\Delta_{\mathrm{ion}}(\bm q)
+\Delta_{\mathrm{xc}}(\bm q)$,  $\Delta_{\mathrm{ion}}(\bm q)$ and $\Delta_{\mathrm{xc}}(\bm q)$ are defined by Eqs.~(\ref{eq:Delta_ion}) and (\ref{eq:Delta_xc}), correspondingly.  We note that $\Delta_{\mathrm{ion}}(\bm q)$ and $\Delta_{\mathrm{xc}}(\bm q)$ follows from the equilibrium KSDFT wavefunctions, without performing TDDFT calculations.

The dynamic correction $\Delta f_{\mathrm{xc}}(\bm q,\omega)$ in Eq.~(\ref{eq:dynamic_xc_correction}) has several advantageous properties. First, it restores the formally consistent first moment of the DSF, Eq.~(\ref{eq:M1_exact}), derived from the density-matrix response formulation of TDDFT with nonlocal XC potentials. Second, it leaves the static density response $\chi(\bm q,0)$ and the static dielectric function $\epsilon(\bm q,0)$ unmodified. This is desirable because nonlocal XC potentials provide a formally consistent description of the equilibrium density within the KS framework \cite{Garrick_PRX_2020}, whereas the inconsistency addressed here arises specifically in the dynamical response. The correction $\Delta f_{\mathrm{xc}}(\bm q,\omega)$ therefore modifies only the latter, without altering the underlying static response. Third, the additive form of the correction makes
$\Delta f_{\mathrm{xc}}(\bm q,\omega)$ directly compatible with any
other static or dynamic XC kernel. Finally, the correction remains well defined in
the optical limit and therefore allows a consistent treatment of quantities such as the dynamic conductivity, as discussed in App.~\ref{sec:optical_correction}.

\subsection{Implications for plasmons}
\label{sec:plasmon_theory}
We recall that the main purpose of pseudopotentials in KSDFT is to avoid treating the tightly bound core electrons explicitly (often referred to as pseudization). Pseudization substantially reduces the numerical cost, particularly in plane-wave calculations, due to a much smaller basis set required to represent the valence wave functions.  For accurate KSDFT calculations, the pseudopotentials must distinguish between different angular-momentum components of the valence wave function, leading naturally to a nonlocal pseudopotential. Pseudization thus separates the valence-electron excitation spectrum from excitations involving core states, whose effects are incorporated into the pseudopotential.

For materials with a well-defined plasmon excitation, such as
simple metals, warm dense matter, and plasmas, the spectral
representation of $\epsilon^{-1}$ in Eq.~(\ref{eq:epsinv_spectral})
provides a direct connection between the plasmon frequency and the
frequency moments of the DSF.  To show that, we consider the long-wavelength regime with $\omega \gg qv_{\rm char}$, in which the plasmon frequency is large compared with the characteristic frequency scale of single-particle excitations.
Here, $v_{\rm char}$ denotes a characteristic electron velocity: in the strongly degenerate limit it is set by the Fermi velocity $v_F$, whereas in the high-temperature limit it is set by the thermal electron velocity \cite{Arista_PRA_1984}. In this regime, the collective mode is well separated from the dominant single-particle excitation continuum, and the inverse dielectric function can, to leading order, be approximated by an undamped single-pole form,
\begin{equation}
\epsilon^{-1}(\bm q,z)
\simeq
1+
\frac{C_1(\bm q)}
{z^2-C_1(\bm q)},
\label{eq:one_pole_general}
\end{equation}
where
\begin{equation}
C_1(\bm q)=\frac{2\omega_{\rm p,0}^2}{q^2}M_{1}(\bm q)
\label{eq:C2}
\end{equation}
is determined by the first frequency moment of the valence-electron DSF.

\begin{figure*}[t]
\centering
\begin{minipage}[c]{0.35\textwidth}
    \centering
    \includegraphics[width=\linewidth]{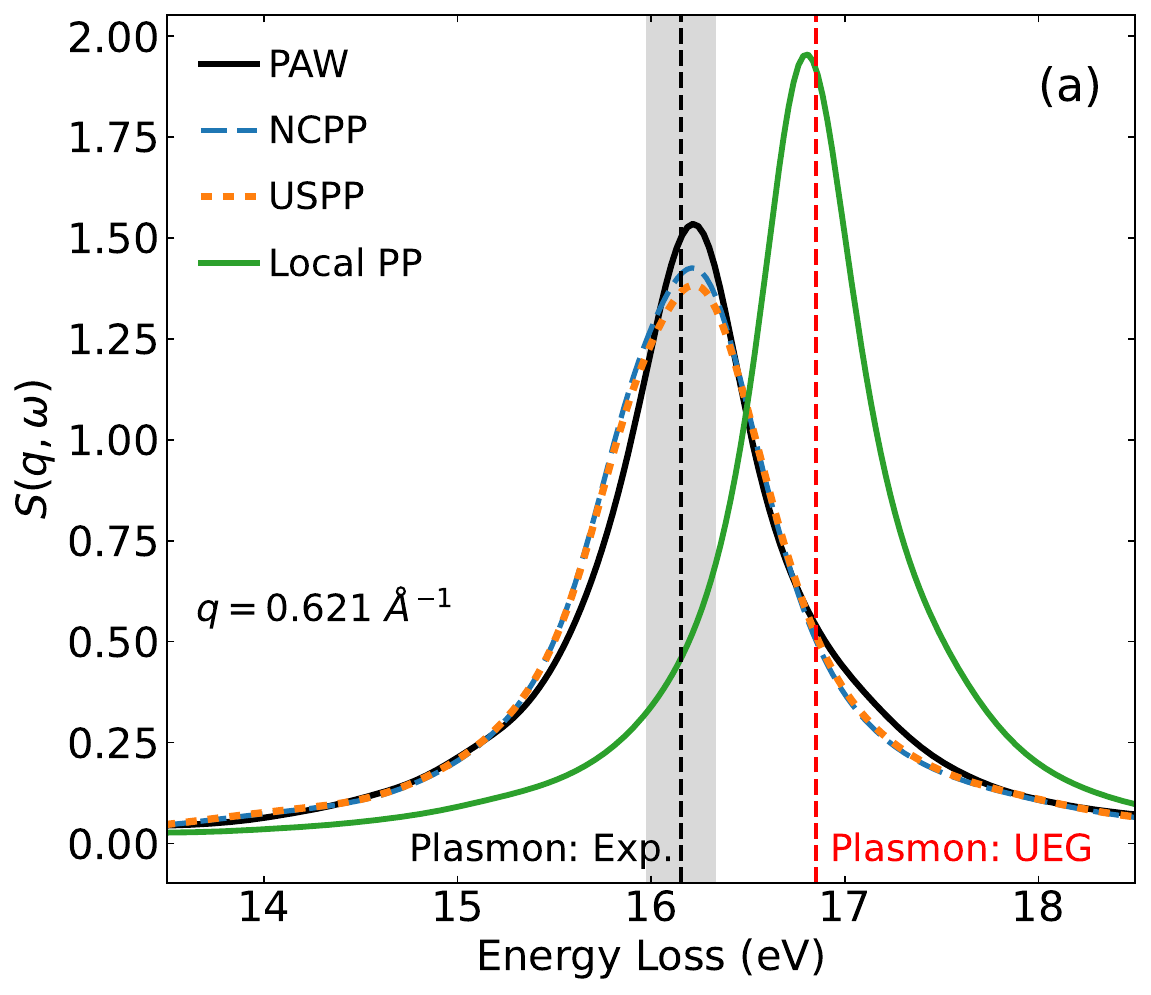}
\end{minipage}
%\hfill
\begin{minipage}[c]{0.6\textwidth}
    \centering
    \includegraphics[width=\linewidth]{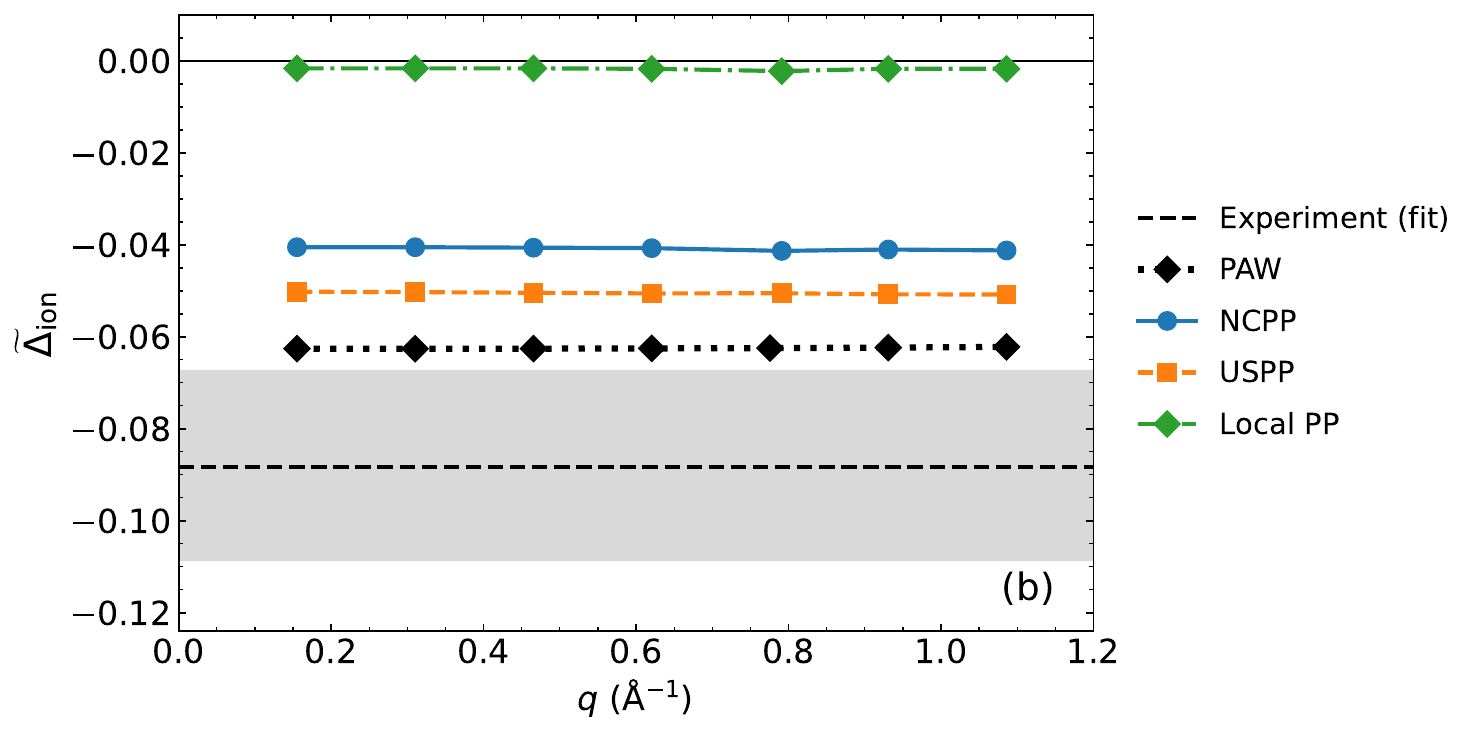}
\end{minipage}

\caption{
Plasmon of ambient aluminum with an fcc crystal structure. (a) TDDFT results for the DSF $S(\mathbf{q},\omega)$ for a fixed wavenumber $q=0.621\,$\AA$^{-1}$ computed from three non-local and one local pseudo potential, see the main text. Vertical lines indicate the plasmon energies obtained from experiment, including the corresponding uncertainty range, and from the uniform electron gas (UEG) model at the same wavenumber. (b) Comparing DFT results for the electron-ion coupling constant $\widetilde{\Delta}_\textnormal{ion}$ [see Eq.~(\ref{eq:Delta_ion_asymptotic})] to the XRTS measurements by Gawne \emph{et al.}~\cite{Gawne_PRB}. The corresponding plasmon dispersion $\omega(q)$ is shown in Fig.~\ref{fig:DSF_Al}(a).
}
\label{fig:Delta_plasmon}
\end{figure*}

The plasmon frequency is determined by a pole of $\epsilon^{-1}(\bm q,z)$. Eq.~(\ref{eq:one_pole_general})
therefore gives, in the long-wavelength limit,
\begin{equation}
\widetilde{\omega}_{\rm p}^2
=\frac{2\omega_{\rm p,0}^2}{q^2}\,M_1(\bm q),
\qquad |\vec q|\ll {\omega}/{v_{\rm char}} .
\label{eq:plasmon_M1}
\end{equation}

Thus, the first frequency moment of the DSF defines the long-wavelength plasmon frequency. 
In general, $\widetilde{\omega}_{\rm p}$ need not coincide with the free-electron plasma frequency 
$\omega_{\rm p,0}=(4\pi n_0)^{1/2}$. The distinction between $\widetilde{\omega}_{\rm p}$ and $\omega_{\rm p,0}$ is already apparent for cold fcc Al. The free-electron plasma frequency associated with the valence-electron density is
$\omega_{\rm p,0}\simeq15.8~{\rm eV}$, whereas experimental XRTS and EELS measurements yield a long-wavelength plasmon energy of approximately
$\widetilde{\omega}_{\rm p}\simeq15~{\rm eV}$
\cite{Gawne_PRB,EELS_PRB_1989}
[see Fig.~\ref{fig:DSF_Al}(a)].
%%Eq.~(\ref{eq:C2}) therefore shows that understanding the TDDFT plasmon energies and the discrepancies in Fig.~\ref{fig:DSF_Al__XRTS} requires a closer examination of the first moment of the DSF, i.e., the $f$-sum rule.

In the limit of small wavenumbers, as shown in  App.~\ref{app:A}, we have 
\begin{equation}
\Delta_{\mathrm{ion}}(\vec q)
=\frac{q^2}{2}\widetilde{\Delta}_{\mathrm{ion}}(\widehat{\bm e}_{\bm q})+o(q^2), \qquad q\rightarrow0,
\label{eq:Delta_ion_asymptotic}
\end{equation}
where $\widehat{\bm e}_{\bm q}$ is the unit vector along $\vec q$ and the factor ${1}/{2}$ is a convention used in this work.

In the case of consistent TDDFT calcuations satisfying Eq.~(\ref{eq:M1_exact}),  substituting $M_1(\bm q)=q^2/2+\Delta_{\mathrm{ion}}(\vec q)$ into Eq.~(\ref{eq:plasmon_M1}), and taking into account Eq.~(\ref{eq:Delta_ion_asymptotic}),  we  derive the result for the plasmon frequency in the long-wavelength limit:
\begin{equation}
\widetilde{\omega}_{\rm p}^2(\widehat{\bm e}_{\bm q})=\omega_{\rm p,0}^2\left(1+\widetilde{\Delta}_{\mathrm{ion}}(\widehat{\bm e}_{\bm q})\right),
\label{eq:plasmon_deltaion_new}
\end{equation}
where, as a rule, the plasmon frequency is generally nearly isotropic,
so that its dependence on the propagation direction
$\widehat{\bm e}_{\bm q}$ can usually be neglected. Typically,
$\widetilde{\Delta}_{\mathrm{ion}}<0$, although its sign is not fixed in
general; a negative value corresponds to a reduction of the plasmon
frequency associated with the nonlocal electron--ion interaction, as we demonstrate it numerically in Sec.~\ref{sec:results1}.

Eq.~(\ref{eq:plasmon_deltaion_new}) has a number of direct implications for
first-principles modeling within KSDFT:
\begin{itemize}
\item Relation~(\ref{eq:plasmon_deltaion_new}) between the plasmon frequency and the correction arising from the nonlocal pseudopotential provides a direct physical interpretation of $\Delta_{\mathrm{ion}}(\bm q)$. Rather than representing a violation of the $f$-sum rule, this term is required for a consistent description of plasmon excitations within the
pseudopotential framework. Its contribution to the first moment in
Eq.~(\ref{eq:M1_exact}) therefore has direct consequences for the predicted excitation spectrum.

%%\item There is no pseudization employing a local pseudopotential that can  reproduce the correct long-wavelength plasmon frequency of a real material, because locality enforces $\Delta_{\mathrm{ion}}(\bm q)=0$ and, consequently, $\widetilde{\omega}_{\rm p}^2=\omega_{\rm p,0}^2$, which is valid only in the nearly free-electron gas limit.

\item For a pseudization in which the chemically inert, tightly bound core electrons are fully incorporated into the pseudopotential, a local pseudopotential cannot reproduce the correct long-wavelength plasmon frequency of the real material. Locality enforces
$\Delta_{\mathrm{ion}}(\bm q)=0$ and, consequently,
$\widetilde{\omega}_{\rm p}^2=\omega_{\rm p,0}^2$, so that the plasmon frequency is determined solely by the explicitly treated valence-electron density. This result is appropriate only in the nearly free-electron-gas limit, where the corresponding ionic nonlocality correction is negligible.

\item Plasmon frequencies measured by XRTS or EELS provide a direct experimental probe of the nonlocal electron--ion contribution
$\Delta_{\mathrm{ion}}(\bm q)$. They therefore offer a physically transparent benchmark for testing whether a pseudopotential captures
the correct plasmon-frequency shift and, more generally, remains transferable in response calculations across different temperatures
and compression regimes.

\item The contribution $\Delta_{\mathrm{xc}}(\bm q)$ to the first moment must be compensated when a nonlocal XC potential is used. If this contribution remains in the first moment of the conventional TDDFT response, it produces an unphysical shift of the plasmon frequency
$\widetilde{\omega}_{\rm p}(\widehat{\bm e}_{\bm q})$. This provides a
direct explanation for the failure of r2SCAN and HSE06 to reproduce the
DSF of fcc aluminum shown in Fig.~\ref{fig:DSF_Al__XRTS}.
\end{itemize}

We demonstrate the application of these findings in Sec.~\ref{sec:applications}.

%%%%%%%%%%============================================
% Figure: Dispersion
% 

\subsection{Plasmons in all-electron simulations}
\label{sec:all_el_f_sum_rule}

The preceding derivations are formulated within the KSDFT description with pseudopotentials. In this framework, the spectral contributions associated with tightly bound core states are excluded from the excitation spectrum through pseudization. Therefore, they do not appear as separate electronic excitations in the dynamic dielectric function given by Eq.~(\ref{eq:epsinv_spectral}). Their influence is instead incorporated indirectly through the
resulting valence-electron structure. This spectral separation makes Eq.~(\ref{eq:plasmon_M1}), derived within
the single-pole approximation, particularly useful for systems with a
well-defined valence-electron plasmon.

In an all-electron description, by contrast, the electron--ion
interaction is represented by the local Coulomb potential and hence
$\Delta_{\mathrm{ion}}(\bm q)=0$. In the absence of a net nonlocal-XC
contribution to the first moment, the familiar all-electron $f$-sum
rule $M_1^{\mathrm{AE}}(\bm q)={q^2}/{2}$ is recovered. However, if the contribution
$\Delta_{\mathrm{xc}}(\bm q)$ due to a non-local XC potential is  not treated consistently, for example through the density-matrix response
formalism or the dynamic XC correction derived above, the calculated
DSF violates the all-electron $f$-sum rule
$M_1^{\mathrm{AE}}(\bm q)=q^2/2$.

For plasmons, we note that in all-electron \textit{ab initio} calculations, 
single-pole approximation (\ref{eq:plasmon_M1}) is generally
insufficient, and one must instead use the full dielectric response,
including transitions involving bound states that would otherwise be
excluded from the explicit excitation spectrum by pseudization.

\section{Applications}
\label{sec:applications}

\subsection{Effect of non-local pseudopotential}
\label{sec:results1}

We start by demonstrating the conclusions of Sec.~\ref{sec:theory} concerning the role of pseudopotentials in the $f$-sum rule and plasmon simulations. In particular, we show that the deviation of the first moment of the DSF from $q^2/2$ caused by pseudization is a physically required consequence of the nonlocal ionic potential entering the pseudized Hamiltonian. As a representative example, we consider the plasmon excitation in fcc aluminum under ambient conditions, for which the collective mode is well defined and the influence of the pseudopotential can be assessed directly through comparison with experimental measurements.

%============================================================
\begin{figure*}[t]
\centering
% ============================================================
% Left column: large panel (a)
% ============================================================
\begin{minipage}[t]{0.37\textwidth}
    \vspace*{4.5mm}
    \centering
    \includegraphics[width=\linewidth]{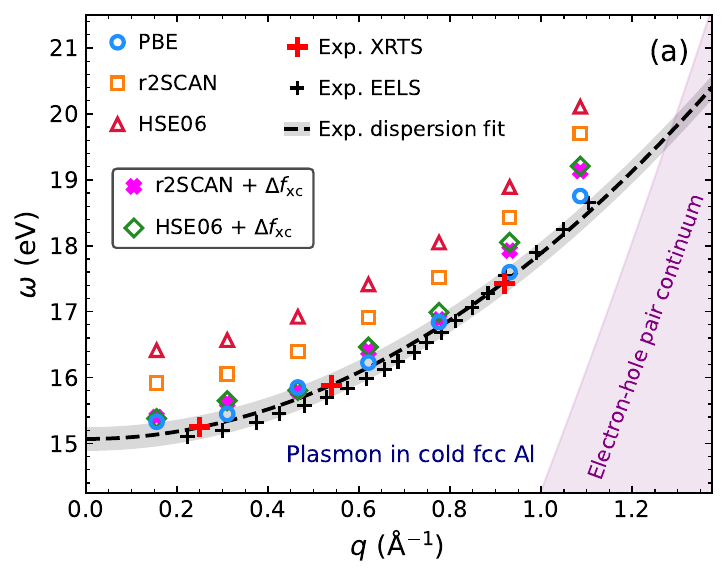}
\end{minipage}
\begin{minipage}[t]{0.355\textwidth}
    \vspace*{0pt}
    \centering
    \includegraphics[width=\linewidth]
    {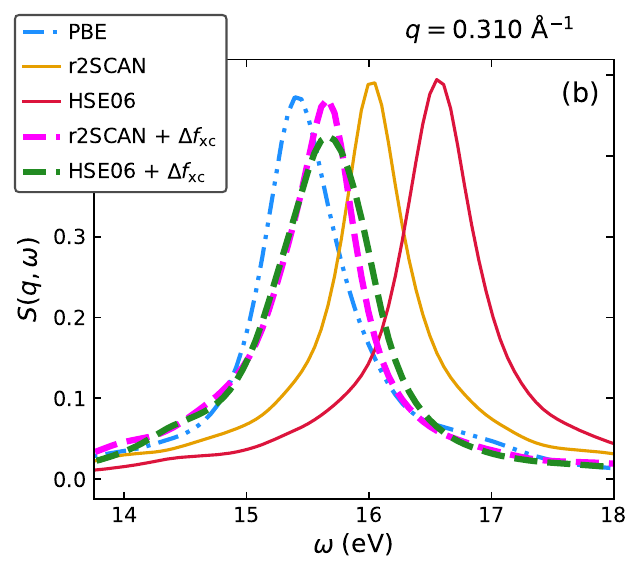}
\end{minipage}
\begin{minipage}[t]{0.219\textwidth}
    \vspace*{4.3mm}
    \centering
    \includegraphics[width=\linewidth]
    {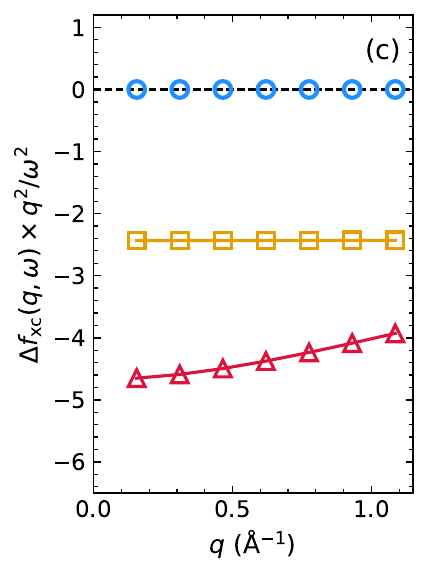}
\end{minipage}

\caption{
(a) Comparison of the plasmon dispersion obtained from TDDFT
calculations using PBE, r2SCAN, and HSE06, including the dynamically
corrected r2SCAN and HSE06 results with XRTS~\cite{Gawne_PRB} and EELS measurements~\cite{EELS_PRB_1989}.
(b) Dynamic structure factor $S(q,\omega)$ at the selected
wave number for the considered XC functionals.
(c) Dynamic exchange--correlation correction for PBE, r2SCAN,
and HSE06.
}
\label{fig:DSF_Al}

\end{figure*}

% ============================================================
% Figure: WDM Al
% ============================================================

Fig.~\ref{fig:Delta_plasmon}(a) shows TDDFT results for the dynamic structure factor of fcc aluminum at $q=0.621~\text{\AA}^{-1}$. For aluminum with three valence electrons per atom, we compare calculations employing three commonly used nonlocal ionic representations---the projector augmented-wave (PAW) \cite{PAW_PRBE_1994, PAW_GPAW}, norm-conserving pseudopotential (NCPP)\cite{NCPP_PRL_1979, Al_NCPP}, and ultrasoft pseudopotential (USPP) \cite{Vanderbilt_PRB, Al_USPP} methods---with a calculation based on a local pseudopotential \cite{local_pp_huang_carter, OFPP_GitHub}. All calculations were performed using the PBE XC functional in combination with the adiabatic LDA (ALDA) XC kernel. More computation details are provided in App.~\ref{app:Comp_det}. 
The three nonlocal approaches yield very similar spectra, with good agreement of plasmon peak positions and line shapes. In contrast, the local pseudopotential produces a pronounced shift of the plasmon peak to higher energy and a substantially different spectral broadening. The corresponding deviations of the first moment $M_1(\bm q)$ from the reference value obtained by neglecting $\Delta_{\mathrm{ion}}$, i.e.,
$M_1(\bm q;\Delta_{\mathrm{ion}}=0)=q^2/2$, are $-6.24\%$, $-4.10\%$, and $-5.07\%$ for PAW, NCPP, and USPP, respectively. The local pseudopotential, on the other hand, gives a deviation of only $-0.17\%$, which lies within the numerical uncertainty associated with the finite frequency range of the calculations ($\omega \leq 100~\mathrm{eV}$) and the use of a Lorentzian broadening of $0.25~\mathrm{eV}$. These differences in $M_1(\bm q)$ compared to $q^2/2$ directly quantify the modification of the spectral weight caused by the nonlocal pseudopotential. In addition, according to Eq.~(\ref{eq:plasmon_deltaion_new}), the same modification of the first moment leads to a corresponding change in the plasmon energy in the long-wavelength limit. For a local pseudopotential, because $\widetilde{\Delta}_{\mathrm{ion}}=0$, the plasmon energy remains close to the value of the free electron gas, see vertical dashed red line in Fig.~\ref{fig:Delta_plasmon}(a). In contrast,  $\Delta_{\mathrm{ion}}$ obtained with PAW, NCPP, and USPP reduces both the first moment and the plasmon energy relative to this reference. This is precisely the behavior observed in Fig.~\ref{fig:Delta_plasmon}(a), where all three nonlocal pseudopotentials yield a comparable downward shift of the plasmon peak relative to the local-pseudopotential result. The simultaneous reduction of the spectral weight and of the plasmon energy produced by the nonlocal pseudopotentials is therefore essential for obtaining good agreement with the experimental DSF measurements; the experimental plasmon position (related uncertainty) is indicated by the vertical dashed black line (shaded grey area). This can also be seen clearly in Fig.~\ref{fig:DSF_Al__XRTS}, where the PBE calculations performed within the PAW framework agrees well with the measured XRTS (with the remaining differences mostly attributed to the finite size of the used detector \cite{Gawne_PRB}). This  demonstrates that the deviation of the first moment of the DSF from $q^2/2$ due to nonlocal character of the pseudopotential has a direct, observable and physically motivated influence on the dynamic response.

Although the single-pole approximation underlying Eq.~(\ref{eq:plasmon_deltaion_new}) neglects the plasmon broadening, it captures the mechanism governing the plasmon-energy shift in the long-wavelengths limit. This shift persists at finite wave numbers as the plasmon follows a Bohm--Gross-type dispersion \cite{Hamann_cpp}. We can therefore exploit the measured plasmon dispersion to infer the effective shift $\widetilde{\Delta}_{\mathrm{ion}}$ associated with the interaction of the valence electrons with the ions, where each ion comprises the nucleus together with the chemically inert, tightly bound core electrons.
To this end, we employ benchmark data from recent XRTS measurements at the European XFEL for aluminum under ambient conditions~\cite{Gawne_PRB}, which were performed in a forward-scattering geometry with sufficient spectral resolution to resolve the plasmon excitations, providing a useful benchmark for theoretical models. In addition, we use the data for plasmon energies inquired via EELS measurements in Ref.~\cite{EELS_PRB_1989}. 
We use a fit to the XRTS measurements of the plasmon frequency in aluminum, $\omega_{\rm exp.}=15.067\pm 0.18~\mathrm{eV}$ ($q$ in units of $\mathrm{\AA}^{-1}$), from Ref.~\cite{Gawne_2025}, with uncertainty estimated by taking into account the measurement resolution as well as by comparing to the EELS data from Ref.~\cite{EELS_PRB_1989}.
Using Eq.~(\ref{eq:plasmon_deltaion_new}), and taking into account that the long-wavelength limit of the plasmon in aluminum can be considered with good accuracy to be direction independent, we get $\widetilde{\Delta}_{\mathrm{ion}}^{\mathrm{exp.}}=\widetilde{\omega}_{\rm p, \rm exp.}^2/{\omega}_{\rm p,0}^2-1$, where $\widetilde{\omega}_{\rm p, \rm exp.}=15.067~\mathrm{eV}$ and ${\omega}_{\rm p,0}=15.78~\mathrm{eV}$.

\begin{figure*}[t]
\centering
\includegraphics[
    width=0.62\textwidth
]{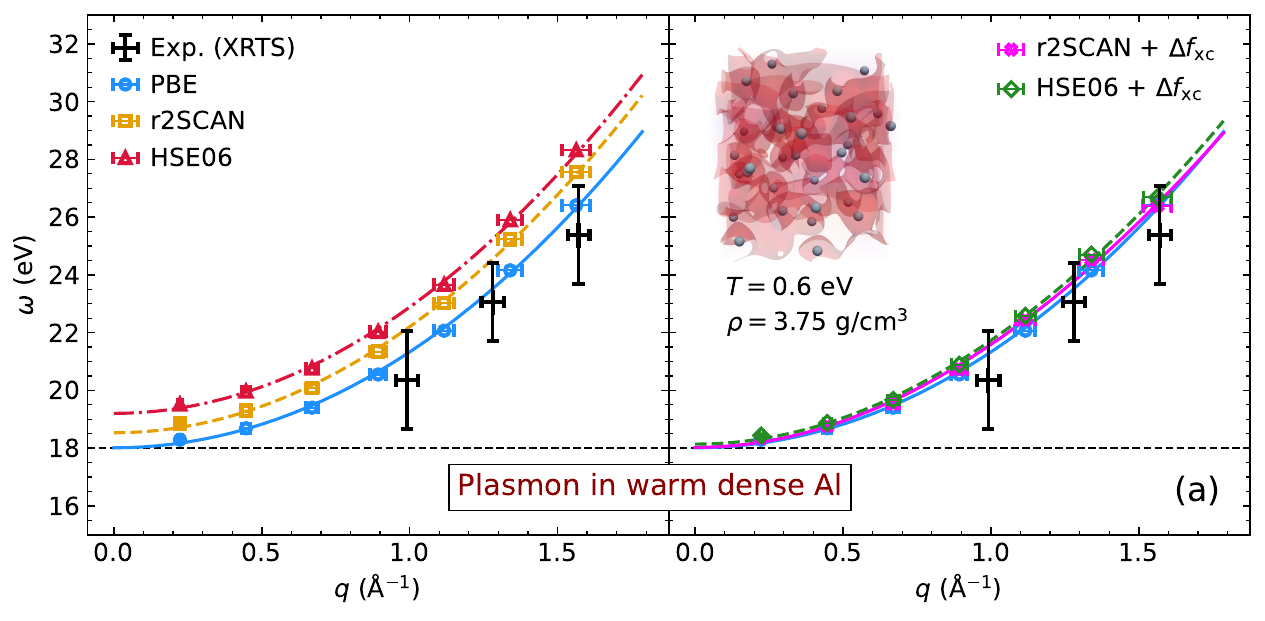}
%\\[-1.mm]
\includegraphics[
    width=0.35\textwidth
]{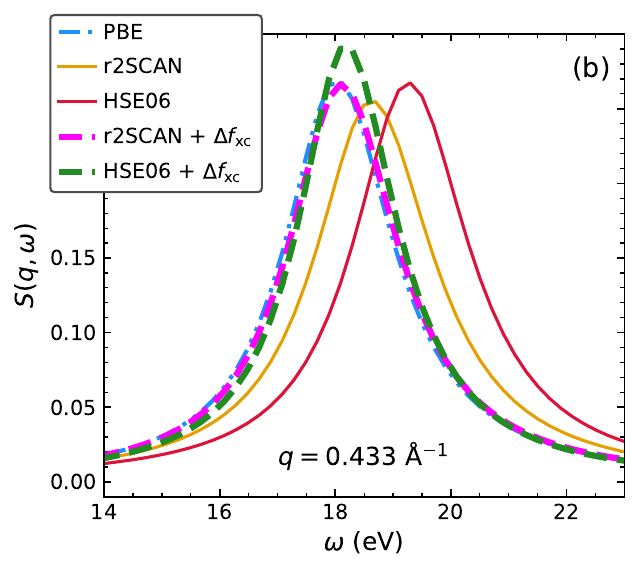}

\caption{
Comparison of TDDFT results with the experimental XRTS signal \cite{BespalovPRL} for
warm dense Al. Panel (a) shows the plasmon dispersion obtained
using PBE, r2SCAN, and HSE06 XC functionals and the corresponding
results including $\Delta f_{\mathrm{xc}}(\mathbf{q},\omega)$. The inset in panel (a) illustrates the electron-density isosurface in a disordered ion snapshot of compressed aluminum at $3.75~\mathrm{g/cm^3}$ and $T=0.6~\mathrm{eV}$.
 Panel (b) shows the dynamic structure factor at density $3.75~\mathrm{g/cm^3}$ and $T=0.6~\mathrm{eV}$.
}
\label{fig:DSF_Al_wdm}
\end{figure*}

Fig.~\ref{fig:Delta_plasmon}(b) shows the $\widetilde{\Delta}_{\mathrm{ion}}$ values computed using different pseudopotentials at finite wavenumbers in the range well below the electron--hole continuum. We compare the KSDFT calculations with $\widetilde{\Delta}_{\mathrm{ion}}^{\mathrm{exp.}}$, inferred by extrapolating the experimental fit to the long-wavelength limit. From Fig.~\ref{fig:Delta_plasmon}(b), we see that $\widetilde{\Delta}_{\mathrm{ion}}$ is nearly independent of $q$, confirming the analytically derived general result of Eq.~(\ref{eq:Delta_ion_asymptotic}). Fig.~\ref{fig:Delta_plasmon}(b) also confirms that the application of a local pseudopotential 
leads to $\widetilde{\Delta}_{\mathrm{ion}}\equiv0$ within numerical accuracy, therefore missing the physically mandated correction to the plasmon frequency of a free electron gas. 
%This deviation is reduced by the use of nonlocal pseudopotentials.
In contrast, all three non-local pseudopotentials correct the plasmon frequency in the right direction.
In particular, the PAW-based calculations show the closest agreement with $\widetilde{\Delta}_{\mathrm{ion}}^{\mathrm{exp.}}$. This is notable because these pseudopotentials are designed primarily for accurate modeling of equilibrium static properties, without explicit consideration of plasmon excitations.

In conclusion, Eq.~(\ref{eq:Delta_ion_asymptotic}) opens an interesting new route for probing electron--ion interaction effects in the regime dominated by collective plasmon excitations and opens up a new avenue for the assessment of advanced, non-local pseudopotentials against accurate XRTS, EELS, or potentially conductivity measurements.

%Here, we reiterate that by an ion we mean a nucleus together with the chemically inert, tightly bound core electrons surrounding it. 

%Beyond purely academic interest, this has practical importance for simulations of compressed and/or heated systems, where it is not always clear in advance what fraction of the electrons should be subjected to pseudization. First, from a computational point of view, this question is particularly important for relatively heavy elements, for which all-electron calculations are challenging. Second, it provides a way to distinguish chemically inert bound electrons from valence electrons under extreme conditions, a problem often identified as one of the challenges in warm dense matter~\cite{vorberger2025roadmapwarmdensematter}.

\subsection{Effect of non-local XC potential}\label{sec:my_name_is_dark}

Among orbital-dependent nonlocal XC functionals, the strongly constrained and appropriately normed (SCAN) functional \cite{SCAN} and the variations of range-separated hybrid Heyd--Scuseria--Ernzerhof (HSE) functionals \cite{HSE03,HSE06,HSE06_Krukau} have played a particularly important role by reducing self-interaction errors while retaining broad transferability. The regularized r2SCAN functional \cite{r2SCAN} was subsequently introduced to alleviate the numerical instabilities of the original SCAN while largely preserving its accuracy and transferability, and has since become widely adopted in practical DFT applications \cite{Riemelmoser2026,Song2023}. 
We use HSE06~\cite{HSE06_Krukau} and r2SCAN~\cite{r2SCAN} as representative examples to demonstrate the $f$-sum-rule inconsistencies that can arise when orbital-dependent XC functionals are employed in the standard TDDFT density response calculations, and to show that these inconsistencies can be readily removed within the density-response framework by applying the proposed dynamic correction to the XC kernel.
All results in this subsection were performed using the PAW method \cite{BlochlPAW}. The parameters of these calculations are provided in more detail in App.~\ref{app:Comp_det}.

Using XRTS and EELS measurements as stringent benchmarks, we first examine ambient aluminum, then extend the analysis to laser-compressed and heated aluminum, and finally turn to ambient diamond silicon and carbon as representative semiconductors. These results demonstrate the versatility of the presented results across markedly different thermodynamic regimes and qualitatively distinct electronic states, ranging from metallic systems to semiconducting phases.

\subsubsection{Ambient aluminum}
\label{sec:appl_fccal}

In Fig.~\ref{fig:DSF_Al}(a), we compare the plasmon dispersion obtained from TDDFT calculations using the r2SCAN, HSE06, and PBE functionals with experimental results from XRTS~\cite{Gawne_PRB} and EELS~\cite{EELS_PRB_1989}. Here, PBE serves as a useful reference, since it is arguably the most widely used GGA XC functional for solids. For each functional, the noninteracting response $\chi_0$, screening, and local-field effects are treated consistently, while the XC kernel is approximated within ALDA.
Fig.~\ref{fig:DSF_Al}(a) shows that the PBE results are in good agreement
with the experimental plasmon dispersion over the considered wavenumber
range, consistent with the previous TDDFT analysis of the XRTS
data~\cite{Gawne_PRB}. Strikingly, both r2SCAN and HSE06 substantially
overestimate the plasmon energy.  For PBE-based linear-response TDDFT, one has $\widetilde{\Delta}_{\mathrm{xc}}(\widehat{\bm e}_{\bm q})=0$,
because the PBE XC potential does not have a non-local part. Consequently, the plasmon frequency that is consistent with Eq.~(\ref{eq:plasmon_deltaion_new}) is recovered. In contrast, for r2SCAN and HSE06 the TDDFT calculations
contain an additional unphysical contribution
$\widetilde{\Delta}_{\mathrm{xc}}(\widehat{\bm e}_{\bm q})$, which
shifts the plasmon frequency away from the physically correct result
and, in the present case, produces a blue shift.

% ============================================================
% Figure: Silicon
% ============================================================

% 

The practical impact of the correction $\Delta f_{\mathrm{xc}}(\bm q,\omega)$ is evident in Fig.~\ref{fig:DSF_Al}(a), where the r2SCAN and HSE06 results including the dynamic XC correction are shown as the purple crosses and green diamonds. In both cases, restoring the $f$-sum-rule consistency shifts the plasmon dispersion into good agreement with experiment, correcting the substantial overestimation obtained from the uncorrected TDDFT response. This demonstrates that the discrepancy is not an inherent failure of the underlying r2SCAN or HSE06 electronic structure, but originates from an inconsistent treatment of the nonlocal XC contribution in the TDDFT response calculation. An example of the DSF obtained with and without the dynamic XC correction is shown in Fig.~\ref{fig:DSF_Al}(b), together with the corresponding PBE result. The latter was also shown to be in good agreement with experiment, within the experimental resolution, at the small wavenumbers considered here~\cite{Gawne_PRB}. The corresponding dynamic XC corrections, shown in the scaled form $\delta f_{\rm xc}^{(1)}(\vec q,\omega) q^2/\omega^2$, are presented in Fig.~\ref{fig:DSF_Al}(c).

\subsubsection{Warm dense aluminum}\label{sec:Al_is_warm}

We next examine the applicability of the dynamic XC correction to warm
dense aluminum generated by laser-induced compression and heating.
Recently, Bespalov \textit{et al.}~\cite{BespalovPRL} reported
high-quality XRTS spectra of warm dense aluminum in the density range
$3.75$--$4.5~\mathrm{g/cm^3}$ and at a temperature
$T\simeq 0.6~\mathrm{eV}$. We performed TDDFT calculations for densities of $3.75$, $4.0$, $4.25$, and $4.5~\mathrm{g/cm^3}$. For each density, ten
independent 32-atom snapshots were selected from separate KSDFT-based
molecular-dynamics simulations, and the final response was averaged over
both snapshots and densities.

In Fig.~\ref{fig:DSF_Al_wdm}(a), we compare the calculated plasmon
dispersion in warm dense aluminum with the experimental XRTS data for
PBE, r2SCAN, and HSE06. Representative DSFs are shown in
Fig.~\ref{fig:DSF_Al_wdm}(b) for
$q=0.433~\mathrm{\AA}^{-1}$. In Fig.~\ref{fig:DSF_Al_wdm}(a),  we also illustrate the ionic structure of compressed aluminum at $3.75~\mathrm{g/cm^3}$ and $T\simeq0.6~\mathrm{eV}$, together with an electron-density isosurface corresponding to the mean electron density $n_0$.

As seen in the left part of Fig.~\ref{fig:DSF_Al_wdm}(a), the PBE results agree with the
measurements within the experimental uncertainty, whereas the
uncorrected r2SCAN and HSE06 calculations systematically overestimate
the plasmon frequency. Applying the dynamic XC correction shifts the
calculated plasmon dispersion toward lower energies and brings both
r2SCAN and HSE06 into agreement with the experimental data, see the right panel of Fig.~\ref{fig:DSF_Al_wdm}(a). This result
is particularly significant because it demonstrates that the same
sum-rule-based dynamic XC correction remains effective across markedly
different thermodynamic regimes, from ambient solids to disordered warm dense
matter.

\subsubsection{Sum-rule consistency for semiconductors}\label{sec:ok}

While hybrid functionals such as HSE06 are generally not recommended for metals, they provide substantial improvements over GGA- and LDA-level XC descriptions for prototypical semiconductors~\cite{Uzulis2025, Schlipf_PRB2011}. Prominent examples are silicon and carbon in the diamond phase at ambient conditions, for which HSE06 provides an accurate description of the direct band gap and yields an improved electronic structure.

\begin{figure}[t]
\centering
\includegraphics[width=\linewidth]{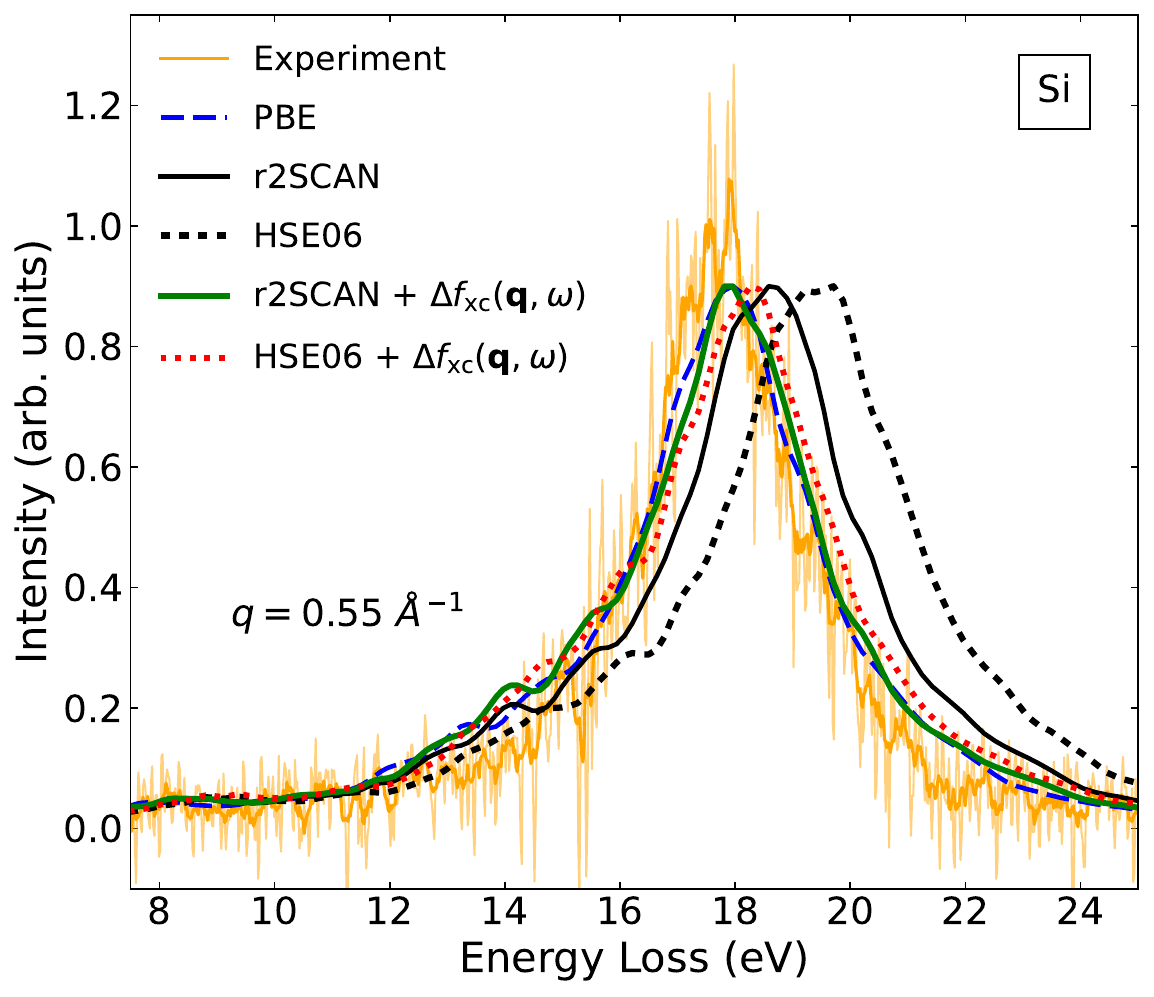}
\caption{
Comparison of the TDDFT results computed using semilocal PBE and
orbital-dependent non-local r2SCAN and HSE06 XC functionals with the experimental
XRTS signal for silicon \cite{Gawne_2025} with a diamond crystal structure.
}
\label{fig:DSF_Si}
\end{figure}

In Fig.~\ref{fig:DSF_Si}, we show the comparison of the TDDFT calculations with the XRTS data from Ref.~\cite{Gawne_2025} for ambient silicon probed at a wavenumber of $q=0.55~\mathrm{\AA}^{-1}$. Specifically, we show PBE, r2SCAN, and HSE06 results, with the latter two shown both with and without the dynamic XC
correction. The PBE result agrees with the XRTS
measurement within the experimental resolution. In contrast, the
uncorrected r2SCAN and HSE06 calculations exhibit a pronounced shift of
the DSF. Applying the dynamic XC correction largely removes
this discrepancy and brings the corresponding TDDFT spectra into substantially improved
agreement with experiment.

\begin{figure}[t]
\centering
\includegraphics[width=\linewidth]{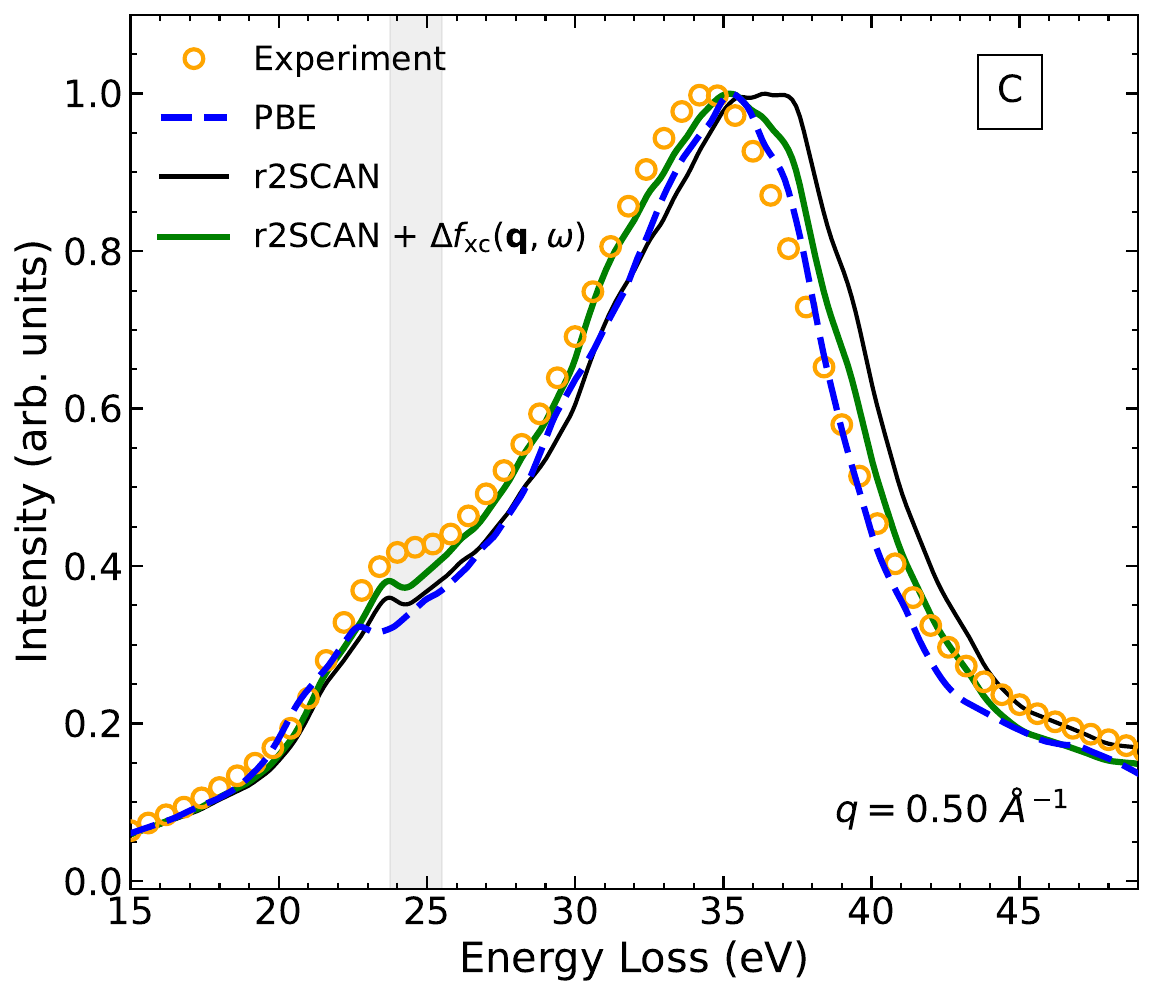}
\caption{
Comparison of the TDDFT results computed using semilocal PBE and
non-local r2SCAN and HSE06 XC functionals with the experimental
XRTS signal for carbon~\cite{Waidmann_EELS_Carbon} with a diamond crystal structure. The shaded grey area indicates a shoulder feature arising from interband transitions.
}
\label{fig:DSF_C}
\end{figure}

As a final example, Fig.~\ref{fig:DSF_C} compares TDDFT results obtained using PBE and r2SCAN with the EELS measurements for diamond carbon from Ref.~\cite{Waidmann_EELS_Carbon}. The TDDFT calculations for diamond carbon require a substantially denser $k$-point grid than in the examples considered above. In combination with the large number of empty bands needed to converge the DSF, this prevented us from obtaining sufficiently converged results with the computationally demanding HSE06 functional.

Compared with the experimental data, Fig.~\ref{fig:DSF_C} shows that the PBE-based results underestimate the width of the plasmon peak and fail to reproduce the pronounced shoulder observed experimentally around $25~{\rm eV}$ (shaded grey area), which is associated with a noncollective interband-transition feature \cite{TIMROV2015460}.
The uncorrected r2SCAN results at least qualitatively capture the shoulder-like feature, but they exhibit the familiar shift of the EELS spectrum toward higher frequencies, leading to a noticeable deviation from the experimental data.

Applying the dynamic correction $\Delta f_{\mathrm{xc}}(\bm q,\omega)$ of Eq.~(\ref{eq:dynamic_xc_correction}) removes this shift, and leads to a noticeably improved agreement in the plasmon width compared to PBE.
Crucially, $\Delta f_{\mathrm{xc}}(\bm q,\omega)$ only corrects the plasmon position, but it does not appear to affect the noncollective interband-transition feature, which continues to appear at the correct position.
The corrected r2SCAN results, thus, show the best agreement with the experimental spectrum.

%Crucially, the r2SCAN results also exhibit a shoulder around $25~{\rm eV}$, in qualitative agreement with the experimental spectrum and in contrast to the PBE results. Together with the removal of the frequency shift by $\Delta f_{\mathrm{xc}}(\bm q,\omega)$, this leads to closer overall agreement of the corrected r2SCAN spectrum with the experimental data.

% Figure: Carbon
%============================================================

\section{Discussion\label{sec:discussion}}

In this work, we have investigated in detail the implications of using non-local pseudopotentials and XC-potentials in linear-response TDDFT calculations of dynamic properties.

First, we have shown that, in lieu of all-electron simulations, non-local pseudopotentials are indispensable to estimate the correct plasmon shift. 
%In contrast, local pseudopotentials always give the uncorrected plasma frequency of a free uniform electron gas. 
In contrast, for pseudizations in which the chemically inert, tightly bound core electrons are entirely incorporated into the pseudopotential, local pseudopotentials always give the uncorrected plasma frequency of a free uniform electron gas.
This result facilitates a new, direct way to benchmark non-local pseudopotentials against XRTS or EELS measurements in the collective regime directly from an equilibrium DFT simulation (i.e., without the need for TDDFT), which is demonstrated here for the case of ambient Al.
In addition, this result has also important implications for orbital-free TDDFT, which are given in more detail below.

Second, we have presented an in-depth investigation of the effects due to non-local XC-potentials, such as meta-GGA and hybrid XC-functionals.
A consistent treatment of the non-locality would require one to take into account the linear response of the full off-diagonal density matrix~\cite{Kuemmel_RevModPhys_2008,  Lacombe2023, Ullrich_Carsten_2025}. 
This would result in a dramatic increase in the needed computational resources that is often prohibitive for practical applications; this problem is further exacerbated for warm dense matter, where the temperature increases the number of orbitals that are required for convergence.
To overcome this bottleneck, we have derived a non-empirical correction $\Delta f_\textnormal{xc}(\mathbf{q},\omega)$ of the dynamic XC-kernel that re-enforces the exact f-sum rule of the non-local KS Hamiltonian. 
This correction can be used for both adiabatic and non-adiabatic kernels, and it can be easily used with standard linear-response TDDFT implementations without any additional computational cost; a corresponding implementation for the open-source \texttt{GPAW} code is made freely available~\cite{GitLab_codes}.

To rigorously demonstrate the value of this new framework, we have compared TDDFT results for GGA (PBE), meta-GGA (r2SCAN) and hybrid (HSE06) XC-functionals for a representative sets of experimental measurements (both XRTS and EELS) of the DSF $S(\mathbf{q},\omega)$ that is comprised of two metallic systems (ambient aluminum and warm dense aluminum) and two prototypical semi-conductors (silicon and carbon with the diamond crystal structure).
Our key findings are: (i) the aforementioned inconsistency in the treatment of non-local XC-potentials in standard TDDFT leads to a poor performance of uncorrected r2SCAN and HSE06 compared to the simpler PBE; (ii) this deficiency is readily corrected by $\Delta f_\textnormal{xc}(\mathbf{q},\omega)$, which leads to a dramatic improvement in all considered cases; (iii) for ambient carbon, $\Delta f_\textnormal{xc}(\mathbf{q},\omega)$ corrects the plasmon feature, but retains the correct position of the prominent interband-transition feature for r2SCAN.

We are confident that our work opens up new avenues for a plethora of impactful future works in a variety of different fields and directions.

\medskip
\textbf{(i)} The possibility to use TDDFT to accurately estimate dynamic properties from orbital-dependent XC-functionals without previous inconsistencies constitutes a mandatory step on the long road towards a truly universal XC-functional that can be used to accurately estimate all observable from a single DFT simulation.
Indeed, it has become well established that hybrid functionals are needed to correctly describe bandgaps in different materials, and the same can be expected for the correct description of interband-transition features in XRTS, EELS or conductivity measurements.
Moreover, hybrid functionals likely have a prominent future in the study of warm dense matter~\cite{bonitz2024principles,vorberger2025roadmapwarmdensematter,Karasiev2026,Witte_PRL,Moldabekov2023}, where TDDFT simulations are in high demand for the interpretation of experiments and to generate data tables for IFE modeling and astrophysics.

\medskip
\textbf{(ii)} A particularly promising route towards improved, non-local XC-functionals is given by the connection of the adiabatic-connection formula with the fluctuation--dissipation theorem (ACFD)~\cite{pribram,Patrick2015}.
Here, the XC-energy is expressed in terms of the dynamic linear density response, either neglecting the dynamic XC-kernel (usually dubbed random phase approximation, RPA) 
or using a suitable approximation, e.g., based on a uniform electron gas~\cite{Patrick2015}.
In either case, this orbital-dependent formulation is afflicted by the same inconsistencies as linear-response TDDFT calculations.
Our correction $\Delta f_\textnormal{xc}(\mathbf{q},\omega)$ would provide a fast and easy fix that avoids considering the substantially more involved full off-diagonal density response, at no additional computational cost.

%In its most widely used random-phase approximation (RPA) form, the XC kernel is neglected, whereas beyond-RPA approaches often employ approximate static kernels \cite{Patrick2015}. In all of these cases, when the input density-response functions are calculated using orbital-dependent nonlocal XC potentials, the resulting $f$-sum-rule inconsistencies and their consequences for dynamic response properties can be straightforwardly corrected using the dynamic kernel derived in this work. Considering the advantages of nonlocal functionals for accurate electronic-structure modeling, together with the negligible additional computational cost of the proposed dynamic correction, the present work provides a valuable practical step toward improving the general predictive capability of KSDFT.

\medskip
\textbf{(iii)} %For linear response TDDFT calculations, we provide the dynamic correction to the XC kernel restoring the $f$-sum rule consistency and related spectral weight and plasmon inconsistencies. We focused on XRTS and EELS calculations, where the correction  $\Delta f_{\mathrm{xc}}(\bm q,\omega)=\Delta f_{\mathrm{xc}, \bm G\bm G}(\bm k,\omega)$ is applied at the level of the diagonal part of the density response matrix $\chi_{\bm G\bm G}(\bm k,\omega)$. 
In the present work, the dynamical correction $\Delta f_{\mathrm{xc}}(\bm q,\omega)=\Delta f_{\mathrm{xc}, \bm G\bm G}(\bm k,\omega)$ is applied at the level of the diagonal part of the density response matrix $\chi_{\bm G\bm G}(\bm k,\omega)$. 
To study the effect of the correction on a deeper level directly entangled with the local field effect,
future work might consider the full matrix form $\Delta f_{\mathrm{xc}, \bm G\bm G^{\prime}}(\bm k,\omega)$, provided in App.~\ref{app:dynamic_kernel_matrix}.

\medskip
\textbf{(iv)} While we have focused here on the DSF for a finite momentum transfer $\hbar q$, the extension of our results to the optical limit of $q\to0$ (often evaluated within the approximate Kubo-Greenwood framework~\cite{wdm_book}) is straightforward; see App.~\ref{sec:optical_correction} for a corresponding derivation and correction.
A second relevant application is given by the correction to the conductivity $f$-sum rule, i.e., the Thomas--Reiche--Kuhn (TRK) sum rule for non-local XC potentials. This correction is particularly important when the conductivity is used to quantify effective ionic charges/ionization states at WDM conditions~\cite{Mandy_PRR_2020}. The TRK sum rule is also widely used as a convergence test for Kubo Greenwood calculations. If the contributions from $\Delta_{\rm ion}$ and $\Delta_{\rm xc}$ are not taken into account, an apparent violation of the sum rule may be misinterpreted as incomplete numerical convergence rather than as a genuine consequence of the nonlocal operators entering the Hamiltonian.

%The implications for the dynamic conductivity in the optical limit, often evaluated within the Kubo--Greenwood (KG) framework~\textcolor{red}{[cite]}, are likewise important. When the underlying electronic structure is obtained using a nonlocal XC potential, the resulting conductivity can exhibit a shifted plasmon feature. This shift can be corrected using the dynamic XC-kernel correction derived in this work, as discussed in App.~\ref{sec:optical_correction}. 

\medskip
\textbf{(v)} The fundamental inability of local pseudopotentials to estimate the correct plasmon shift it particularly relevant for orbital-free methods \cite{Michele_Chem_Rev,Wenhui_OFDFT}, where local pseudopotentials are commonly employed. Orbital-free TDDFT (OF-TDDFT) is considered a particularly promising route for extending TDDFT simulations to large systems \cite{PhysRevX.11.011049}. In WDM and IFE applications, one of its important use cases is the calculation of stopping power \cite{Nichols_PRE_2026, Jiang_PRB,Collins_PRL_2018,Malko2022}, where plasmon excitations constitute a major channel of energy dissipation \cite{Arista_PRB_1978,Moldabekov_PRE_2020}. The present results therefore not only identify a fundamental limitation of local pseudopotentials in describing plasmons, but also provide a practical route to overcome it.  Following an approach analogous to that used in Sec.~\ref{subsec:non_imp_xc} to derive the dynamic correction to the XC kernel, the missing $\Delta_{\rm ion}$ contribution to the first moment, and consequently to the plasmon dispersion, can be incorporated into OF-TDDFT through the effective non-local potential
\begin{equation}
\nabla^2 V_{\mathrm{ei}}^{\rm NL}(\bm r,t)
=
\frac{\widetilde{\Delta}_{\mathrm{ei}}}
{n_0\left[1/2+\widetilde{\Delta}_{\mathrm{ei}}\right]}
\frac{\partial^2 n(\bm r,t)}{\partial t^2}.
\label{eq:app_Vei_realspace}
\end{equation}
The corresponding derivation and reciprocal-space representation are provided in App.~\ref{app:OF_TDDFT_dynamic_ei}.

\medskip
\textbf{(vi)} Beyond the specific results presented above, the general theoretical framework developed in this work is itself of broader value. By establishing a direct connection between the high-frequency expansion of linear density response functions, the moments of the DSF, and the Dyson equation for the one-particle density matrix response, it provides a systematic basis for the further development of dynamic XC-kernel corrections and for identifying additional general constraints on nonlocal XC potentials, e.g., by considering the third moment of the DSF.

\section*{Acknowledgements}

This work has received funding from the European Research Council
(ERC) under the European Union's Horizon 2022 research and innovation
programme (Grant agreement No.~101076233, ``PREXTREME'').
Views and opinions expressed are, however, those of the authors only
and do not necessarily reflect those of the European Union or the
European Research Council Executive Agency. Neither the European
Union nor the granting authority can be held responsible for them.
TD gratefully acknowledges funding from the Deutsche Forschungsgemeinschaft (DFG) via project DO 2670/1-1.
MP acknowledges a grant from the US Department of Energy, grant number DE-SC0024496.

Computations were performed on a Bull Cluster at the Center for
Information Services and High-Performance Computing (ZIH) at
Technische Universit\"at Dresden, at the Norddeutscher Verbund
f\"ur Hoch- und H\"ochstleistungsrechnen (HLRN) under grant
mvp00024, and on the high-performance computer Otus at the NHR Center "Paderborn Center for Parallel Computing (PC2)" (project ID 27589, HiFi-WDM).

\section*{Data Availability}

The data supporting the findings of this study will be made available through the Rossendorf Data Repository (RODARE) upon publication~\cite{moldabekov_zhandos_2026_4513}.

\section*{Code Availability}

The open-source code to compute $\Delta_{\mathrm{ion}}(\bm q)$ according to Eq.~(\ref{eq:Delta_ion})  within the PAW method and $\Delta_{\mathrm{xc}}(\bm q)$ using Eq.~(\ref{eq:Delta_xc}) for range-separated hybrid functionals of the HSE type, global hybrid functionals of the PBE0 type, and the meta-GGA functionals r2SCAN and SCAN freely available at~\cite{GitLab_codes}.
% ============================================================
% Acknowledgements
\appendix

\section{Computational details}\label{app:Comp_det}
The PAW-based calculations were performed using the open-source GPAW code~\cite{GPAW_2024,Enkovaara_2010,Mortensen_PRB_2005,GPAW_lrTDDFT}. The calculations were carried out using the PBE, r2SCAN, and HSE06 functionals.
We employed  PBE PAW setups for aluminum, silicon, and carbon, provided with GPAW-PAW-data base. For aluminum at both ambinet and warm dense matter conditions, we used Monkhorst-Pack k-point grid. For both silicon and carbon, we used Gamma-centered Monkhorst-Pack k-point grid.

For ambient fcc aluminum, we employed a four-atom cubic cell with lattice constant $4.05~{\rm \AA}$, a plane-wave cutoff of $350~\mathrm{eV}$, an $8\times8\times8$ $k$-point grid, and 40 bands. The DSF calculations were performed with the wave vector along the crystallographic $[001]$ direction.
Warm dense aluminum was considered at mass densities of $3.75$, $4.0$, $4.25$, and $4.5~\mathrm{g/cm^3}$. For each density, ten independent 32-atom configurations were selected from separate KSDFT molecular-dynamics simulations, and the final spectra were averaged over both configurations and densities. The electronic and ionic temperatures were set to $0.6~\mathrm{eV}$. A plane-wave cutoff of $350~\mathrm{eV}$, a $4\times4\times4$ $k$-point grid, and 400 bands were used, with the large number of bands required to describe thermally populated and excited states.

 The NCPP, USPP, and local-pseudopotential results shown in Fig.~\ref{fig:Delta_plasmon} were obtained using the open-source Quantum ESPRESSO package~\cite{Giannozzi_2017,Giannozzi_2020,Giannozzi_2009,TIMROV2015460}. The local pseudopotential requires a substantially higher plane-wave cutoff and denser $k$-point sampling. We therefore used a cutoff of $1020~\mathrm{eV}$ ($75~\mathrm{Ry}$) and a $20\times20\times20$ $k$-point grid. For a consistent comparison, the same cutoff and $k$-point grid were also used for the NCPP, USPP, and PAW calculations.
 Al.pbe-rrkj.UPF (for NCPP) and Al.pbe-nl-rrkjus psl.1.0.0.UPF (for USPP) from  Quantum ESPRESSO  pseudopotential database were used. A local pseudopotential developed by Huang and Carter \cite{local_pp_huang_carter}, which is available at \cite{OFPP_GitHub}, was employed. All pseudopotentials have three valence electrons per aluminum atom.
 
For diamond silicon, we used an eight-atom cubic cell with lattice constant $5.431~{\rm \AA}$, a $3\times3\times9$ $k$-point grid, a plane-wave cutoff of $350~\mathrm{eV}$, and 100 bands. 
For diamond silicon,  the wave vector was set along the crystallographic $[331]$ direction. 
For diamond carbon, an eight-atom cubic cell with lattice constant $3.567~{\rm \AA}$, a $24\times24\times24$ $k$-point grid, a plane-wave cutoff of $650~\mathrm{eV}$, and 100 bands were employed. The DSF calculations for diamond carbon were performed with the wave vector set along the crystallographic $[111]$ direction.  In all linear-response TDDFT calculations, the dielectric-matrix cutoff was set to $120~\mathrm{eV}$ and a Lorentzian broadening of $0.25~\mathrm{eV}$ was used.
 
The results for $\Delta_{\mathrm{ion}}(\bm q)$ and $\Delta_{\mathrm{xc}}(\bm q)$ were computed using the same simulation parameters as in the corresponding TDDFT calculations. Open-source implementations for evaluating these quantities are provided at~\cite{GitLab_codes}, together with detailed documentation and user instructions.

To facilitate reproducibility, all input files used in the KSDFT and TDDFT calculations will be made available through the Rossendorf Data Repository (RODARE) upon publication~\cite{moldabekov_zhandos_2026_4513}.

\section{Derivation of the density-matrix Dyson equation}
\label{app:density_matrix_dyson}

We briefly derive the Dyson equation for the one-particle
density-matrix response. All quantities below are understood at the
same frequency $\omega$, which is suppressed for compactness.

We use compact numerical labels for one-particle coordinates,
\begin{equation}
1\equiv(\bm r_1,\sigma_1),\qquad
2\equiv(\bm r_2,\sigma_2),\qquad \ldots,
\end{equation}
and adopt the notation
\begin{align}
\gamma(12)&\equiv\gamma(1,2), \\
L(12;34)&\equiv L(1,2;3,4), \\
\mathcal K(12;34)&\equiv\mathcal K(1,2;3,4).
\end{align}
Here, the semicolon in a four-point quantity such as $L(12;34)$
separates the pair of coordinates $(1,2)$ associated with the
density-matrix variation from the pair $(3,4)$ associated with the
perturbation. Integration over a numerical coordinate includes both
the spatial coordinate and, when applicable, the spin variable,
\begin{equation}
\int d1
\equiv
\sum_{\sigma_1}\int d\bm r_1.
\end{equation}

We use the symbol $\circ$ to denote integration over intermediate
one-particle coordinates. For two four-point quantities,
\begin{equation}
(A\circ B)(12;34)
=
\int d5\,d6\,
A(12;56)B(56;34),
\label{eq:app_contraction}
\end{equation}
whereas the action of a four-point kernel on a two-point quantity is
\begin{equation}
(A\circ b)(12)
=
\int d3\,d4\,
A(12;34)b(34).
\label{eq:app_kernel_action}
\end{equation}

Consider a small variation of the equilibrium KS one-particle
Hamiltonian. The kinetic-energy operator and all fixed one-particle
terms are unchanged by the perturbation, so that only the externally
applied field and the self-consistently induced electronic potential
contribute to the first-order variation,
\begin{equation}
\delta\hat h_{\mathrm{GKS}}
=\delta\hat u_{\mathrm{ext}} +
\delta\hat v_{\mathrm{ind}}.
\label{eq:app_dh}
\end{equation}

The noninteracting density-matrix response function $L_0$ is defined
through the first-order relation
\begin{equation}
\delta\gamma(12)
=
\int d3\,d4\,
L_0(12;34)\,
\delta h_{\mathrm{GKS}}(34),
\label{eq:app_L0_response}
\end{equation}
or, in the compact notation introduced above,
\begin{equation}
\delta\gamma
=
L_0\circ\delta h_{\mathrm{GKS}}.
\end{equation}
Using Eq.~(\ref{eq:app_dh}), the density-matrix variation becomes
\begin{equation}
\delta\gamma
=
L_0\circ\delta u_{\mathrm{ext}}
+
L_0\circ\delta v_{\mathrm{ind}}.
\label{eq:app_dgamma_L0}
\end{equation}

The induced one-particle potential changes in response to the
density-matrix variation. To linear order,
\begin{equation}
\delta v_{\mathrm{ind}}(12)
=
\int d3\,d4\,
\mathcal K(12;34)\,
\delta\gamma(34),
\label{eq:app_Vind_explicit}
\end{equation}
where
\begin{equation}
\mathcal K(12;34)
=
\frac{\delta v_{\mathrm{ind}}(12)}
{\delta\gamma(34)}
\end{equation}
is the density-matrix kernel defined in
Eq.~(\ref{eq:K_decomposition}). In compact notation,
\begin{equation}
\delta v_{\mathrm{ind}}
=
\mathcal K\circ\delta\gamma.
\label{eq:app_Vind}
\end{equation}

Substitution of Eq.~(\ref{eq:app_Vind}) into
Eq.~(\ref{eq:app_dgamma_L0}) gives
\begin{equation}
\delta\gamma
=
L_0\circ\delta u_{\mathrm{ext}}
+
L_0\circ\mathcal K\circ\delta\gamma.
\label{eq:app_dgamma}
\end{equation}
This equation expresses the self-consistent character of the
response: the external perturbation produces a density-matrix
variation, which generates an induced potential, and the induced
potential in turn produces an additional density-matrix variation.

The full density-matrix response function $L$ is defined through
\begin{equation}
\delta\gamma(12)
=
\int d3\,d4\,
L(12;34)\,
\delta u_{\mathrm{ext}}(34),
\label{eq:app_L_def_explicit}
\end{equation}
or equivalently,
\begin{equation}
\delta\gamma
=
L\circ\delta u_{\mathrm{ext}}.
\label{eq:app_L_def}
\end{equation}
Substituting Eq.~(\ref{eq:app_L_def}) into
Eq.~(\ref{eq:app_dgamma}) yields
\begin{equation}
L\circ\delta u_{\mathrm{ext}}
=
L_0\circ\delta u_{\mathrm{ext}}
+
L_0\circ\mathcal K\circ L
\circ\delta u_{\mathrm{ext}}.
\label{eq:app_before_Dyson}
\end{equation}

The response functions $L$ and $L_0$ are formally defined with
respect to an arbitrary one-particle perturbation
$\delta u_{\mathrm{ext}}(34)$. Since
Eq.~(\ref{eq:app_before_Dyson}) must hold for an arbitrary
$\delta u_{\mathrm{ext}}$, the kernels acting on this perturbation
must be equal. Therefore,
\begin{align}
L(12;34)=&
L_0(12;34)\nonumber\\
&+\int d5\,d6\,d7\,d8\, L_0(12;56)\,
\mathcal K(56;78)\, L(78;34).
\label{eq:app_L_Dyson}
\end{align}

Eq.~(\ref{eq:app_L_Dyson}) is the density-matrix Dyson equation
used in Eq.~(\ref{eq:L_Dyson}) of the main text.

For the physical response to a local external scalar potential, the
general one-particle perturbation is restricted to
\begin{equation}
\delta u_{\mathrm{ext}}(12)
=
\delta(1-2)\,
\delta v_{\mathrm{ext}}(1),
\label{eq:app_local_source}
\end{equation}
where $\delta(1-2)$ denotes the delta function in the corresponding
one-particle variables. The introduction of an arbitrary two-point
source $\delta u_{\mathrm{ext}}(12)$ is therefore a convenient way of
defining the full four-point density-matrix response function. The
physical external perturbation relevant for the density response
remains a local scalar potential.

\section{High-frequency expansion of the noninteracting density-matrix response}
\label{app:L0_high_frequency}

The one-particle density-matrix operator is
\begin{equation}
\hat\gamma(\bm r,\bm r')
=
\hat\psi^\dagger(\bm r')\hat\psi(\bm r).
\end{equation}
For a noninteracting KS system with one-particle eigenstates
$\varphi_a(\bm r)$, eigenenergies $\epsilon_a$, and equilibrium
occupations $f_a$, the analytically continued density-matrix response is
\begin{align}
{L}_0(\bm r,\bm r';\bm x,\bm x';z)
={}&
\sum_{ab}
\frac{f_a-f_b}
{z-(\epsilon_b-\epsilon_a)}
\nonumber\\
&\times
\varphi_b(\bm r)\varphi_a^*(\bm r')
\varphi_b^*(\bm x)\varphi_a(\bm x').
\label{eq:L0_spectral_app}
\end{align}
Here, $a$ and $b$ denote complete one-particle quantum numbers,
including the band, crystal momentum, and spin where appropriate.

In the asymptotic region of the upper complex-frequency plane,
$|z|\rightarrow\infty$ with ${\rm Im}\,z>0$ while remaining separated
from the real-frequency axis, the spectral denominator can be expanded as
\begin{align}
\frac{1}{z-(\epsilon_b-\epsilon_a)}
={}&
\frac{1}{z}
+
\frac{\epsilon_b-\epsilon_a}{z^2}
+
\mathcal O(z^{-3}).
\label{eq:denominator_expansion_app}
\end{align}
Substitution into Eq.~(\ref{eq:L0_spectral_app}) gives
\begin{equation}
{L}_0(z)
=
\frac{L_0^{(1)}}{z}
+
\frac{L_0^{(2)}}{z^2}
+
\mathcal O(z^{-3}),
\label{eq:L0_expansion_app}
\end{equation}
where
\begin{align}
L_0^{(1)}(\bm r,\bm r';\bm x,\bm x')
={}&
\sum_{ab}(f_a-f_b)
\varphi_b(\bm r)\varphi_a^*(\bm r')
\nonumber\\
&\times
\varphi_b^*(\bm x)\varphi_a(\bm x'),
\label{eq:L01_app}
\end{align}
and
\begin{align}
L_0^{(2)}(\bm r,\bm r';\bm x,\bm x')
={}&
\sum_{ab}(f_a-f_b)(\epsilon_b-\epsilon_a)
\nonumber\\
&\times
\varphi_b(\bm r)\varphi_a^*(\bm r')
\varphi_b^*(\bm x)\varphi_a(\bm x').
\label{eq:L02_app}
\end{align}
Thus, in contrast to the density response, the unprojected
density-matrix response generally contains a nonvanishing term of
order $1/z$.

We next show that this term vanishes upon projection onto the density
response. Applying Eq.~(\ref{eq:Pq_definition}) to
Eq.~(\ref{eq:L01_app}) gives
\begin{equation}
\mathcal P_{\bm q}[L_0^{(1)}]
=
\frac{1}{\Omega}
\sum_{ab}
(f_a-f_b)
\left|
\langle a|e^{-i\bm q\cdot\bm r}|b\rangle
\right|^2 .
\label{eq:PL01_app}
\end{equation}
Separating the two occupation terms,
\begin{align}
\mathcal P_{\bm q}[L_0^{(1)}]
={}&
\frac{1}{\Omega}
\sum_a f_a
\sum_b
\left|
\langle a|e^{-i\bm q\cdot\bm r}|b\rangle
\right|^2
\nonumber\\
&-
\frac{1}{\Omega}
\sum_b f_b
\sum_a
\left|
\langle a|e^{-i\bm q\cdot\bm r}|b\rangle
\right|^2 .
\label{eq:PL01_split_app}
\end{align}
Completeness of the one-particle states and the unitarity of
$e^{-i\bm q\cdot\bm r}$ give
\begin{equation}
\sum_b
\left|
\langle a|e^{-i\bm q\cdot\bm r}|b\rangle
\right|^2
=
1,
\qquad
\sum_a
\left|
\langle a|e^{-i\bm q\cdot\bm r}|b\rangle
\right|^2
=
1.
\end{equation}
Hence,
\begin{equation}
\mathcal P_{\bm q}[L_0^{(1)}]
=
\frac{1}{\Omega}
\left(
\sum_a f_a-\sum_b f_b
\right)
=0.
\label{eq:PL01_zero_app}
\end{equation}
Although ${L}_0$ therefore starts at order $1/z$, its density
projection starts at order $1/z^2$,
\begin{equation}
{\chi}_0(\bm q,z)
=
\mathcal P_{\bm q}[{L}_0](z)
=
\frac{\mathcal P_{\bm q}[L_0^{(2)}]}{z^2}
+
\mathcal O(z^{-3}).
\label{eq:chi0_from_L0_app}
\end{equation}
The symmetry of the density-response spectral representation eliminates
the odd inverse powers, giving
\begin{equation}
{\chi}_0(\bm q,z)
=
\frac{2n_0M_1^{(0)}(\bm q)}{z^2}
+
\mathcal O(z^{-4}),
\label{eq:chi0_M1_app}
\end{equation}
in agreement with Eq.~(\ref{eq:chi_moment_expansion_z}) for the
noninteracting response.

\section{Cancellation of the nonlocal XC contribution from local
gauge covariance}
\label{app:xc_gauge_cancellation}

Here we show that a density-matrix XC kernel consistent with the
nonlocal XC potential exactly cancels the contribution $\Delta_{\mathrm{xc}}(\bm q)$ to the first moment if local gauge covariance is not violated.

We denote the equilibrium one-particle density matrix by
$\hat\gamma_{\rm eq}$, normalized as $\mathrm{Tr}\,\hat\gamma_{\rm eq}=N_e$,
and introduce the one-particle density operator
\begin{equation}
\hat u_{\bm q}=e^{-i\bm q\cdot\hat{\bm r}},
\qquad
\hat u_{-\bm q}=\hat u_{\bm q}^{\dagger}.
\label{eq:uq_app}
\end{equation}
The many-particle density operator is then
$\hat\rho_{\bm q}=\sum_i\hat u_{\bm q}^{(i)}$.

For an arbitrary one-particle perturbation $\delta\hat u$, the
high-frequency expansion of the noninteracting density-matrix
response derived in Sec.~\ref{app:L0_high_frequency} implies
\begin{equation}
L_0^{(1)}\circ\delta\hat u
=
[\delta\hat u,\hat\gamma_0].
\label{eq:L01_commutator}
\end{equation}
Indeed, the linearized equation of motion has the high-frequency form
\begin{equation}
z\,\delta\hat\gamma
=
[\hat h_0,\delta\hat\gamma]
+
[\delta\hat u,\hat\gamma_0],
\end{equation}
and therefore
\begin{equation}
\delta\hat\gamma
=
\frac{[\delta\hat u,\hat\gamma_0]}{z}
+\mathcal O(z^{-2}),
\end{equation}
which proves Eq.~(\ref{eq:L01_commutator}).

For later use, the projection $\mathcal P_{\bm q}$ can equivalently
be written in operator form as
\begin{equation}
\mathcal P_{\bm q}[A]
=
\frac{1}{\Omega}
\mathrm{Tr}
\left\{
\hat u_{\bm q}
\left(A\circ\hat u_{-\bm q}\right)
\right\},
\label{eq:Pq_operator_app}
\end{equation}
where $A$ denotes a four-point response kernel. This is identical to
the real-space definition in Eq.~(\ref{eq:Pq_definition}).

We now consider the nonlocal XC potential as a functional of the
one-particle density matrix,
\begin{equation}
\hat v_{\mathrm{xc}}^{\mathrm{NL}}
=
\hat v_{\mathrm{xc}}^{\mathrm{NL}}[\hat\gamma].
\end{equation}
Under a local gauge transformation
\begin{equation}
\hat U_{\phi}=e^{i\hat\phi},
\qquad
\hat\phi=\phi(\hat{\bm r}),
\end{equation}
the one-particle density matrix transforms as
\begin{equation}
\hat\gamma_{\phi}
=
\hat U_{\phi}\hat\gamma_0\hat U_{\phi}^{\dagger}.
\label{eq:gamma_gauge}
\end{equation}
Local gauge covariance of the nonlocal XC potential requires
\begin{equation}
\hat v_{\mathrm{xc}}^{\mathrm{NL}}[\hat\gamma_{\phi}]
=
\hat U_{\phi}
\hat v_{\mathrm{xc}}^{\mathrm{NL}}[\hat\gamma_0]
\hat U_{\phi}^{\dagger}.
\label{eq:vxc_gauge}
\end{equation}

For an infinitesimal transformation, Eqs.~(\ref{eq:gamma_gauge})
and~(\ref{eq:vxc_gauge}) give
\begin{equation}
\delta\hat\gamma_{\phi}
=
i[\hat\phi,\hat\gamma_0],
\qquad
\delta\hat v_{\mathrm{xc}}^{\mathrm{NL}}
=
i[\hat\phi,\hat v_{\mathrm{xc}}^{\mathrm{NL}}].
\label{eq:gauge_variations}
\end{equation}
On the other hand, by definition of the density-matrix XC kernel,
\begin{equation}
\delta\hat v_{\mathrm{xc}}^{\mathrm{NL}}
=
\mathcal K_{\mathrm{xc}}^{(0)}
\circ\delta\hat\gamma ,
\label{eq:Kxc_definition_app}
\end{equation}
where only the high-frequency limit of the kernel is relevant for the
first moment. Combining Eqs.~(\ref{eq:gauge_variations})
and~(\ref{eq:Kxc_definition_app}) yields the gauge-covariance identity
\begin{equation}
\mathcal K_{\mathrm{xc}}^{(0)}
\circ[\hat\phi,\hat\gamma_0]
=
[\hat\phi,\hat v_{\mathrm{xc}}^{\mathrm{NL}}].
\label{eq:Kxc_gauge_identity}
\end{equation}
Although $\phi(\bm r)$ is real for a physical gauge transformation,
Eq.~(\ref{eq:Kxc_gauge_identity}) is linear in $\phi$ and therefore
holds separately for its Fourier components. In particular,
\begin{equation}
\mathcal K_{\mathrm{xc}}^{(0)}
\circ[\hat u_{-\bm q},\hat\gamma_0]
=
[\hat u_{-\bm q},\hat v_{\mathrm{xc}}^{\mathrm{NL}}].
\label{eq:Kxc_gauge_q}
\end{equation}

We can now evaluate the first kernel contribution to the first moment.
From Eq.~(\ref{eq:DeltaM1_kernel}),
\begin{equation}
\Delta M_1(\bm q)
=
\frac{1}{2n_0}
\mathcal P_{\bm q}
\left[
L_0^{(1)}
\circ\mathcal K_{\mathrm{xc}}^{(0)}
\circ L_0^{(1)}
\right].
\label{eq:DeltaM1_start_app}
\end{equation}
Using Eqs.~(\ref{eq:L01_commutator}) and
(\ref{eq:Pq_operator_app}), and $n_0\Omega=N_e$, we obtain
\begin{align}
\Delta M_1(\bm q)
={}&
\frac{1}{2N_e}
\mathrm{Tr}
\left\{
\hat u_{\bm q}
\left[
\mathcal K_{\mathrm{xc}}^{(0)}
\circ[\hat u_{-\bm q},\hat\gamma_0],
\hat\gamma_0
\right]
\right\}.
\label{eq:DeltaM1_trace1}
\end{align}
Applying the gauge-covariance identity
(\ref{eq:Kxc_gauge_q}) gives
\begin{align}
\Delta M_1(\bm q)
={}&
\frac{1}{2N_e}
\mathrm{Tr}
\left\{
\hat u_{\bm q}
\left[
[\hat u_{-\bm q},
\hat v_{\mathrm{xc}}^{\mathrm{NL}}],
\hat\gamma_0
\right]
\right\}.
\label{eq:DeltaM1_trace2}
\end{align}
Using cyclic invariance of the trace,
\begin{equation}
\mathrm{Tr}\{A[B,\gamma_0]\}
=
\mathrm{Tr}\{\gamma_0[A,B]\},
\label{eq:trace_identity_app}
\end{equation}
we find
\begin{align}
\Delta M_1(\bm q)
={}&
\frac{1}{2N_e}
\mathrm{Tr}
\left\{
\hat\gamma_0
[\hat u_{\bm q},
[\hat u_{-\bm q},
\hat v_{\mathrm{xc}}^{\mathrm{NL}}]]
\right\}.
\label{eq:DeltaM1_trace3}
\end{align}
Since
$[\hat u_{\bm q},\hat u_{-\bm q}]=0$, the Jacobi identity gives
\begin{equation}
[\hat u_{\bm q},
[\hat u_{-\bm q},\hat v_{\mathrm{xc}}^{\mathrm{NL}}]]
=
-
[\hat u_{-\bm q},
[\hat v_{\mathrm{xc}}^{\mathrm{NL}},\hat u_{\bm q}]].
\label{eq:Jacobi_app}
\end{equation}
Consequently,
\begin{equation}
\Delta M_1(\bm q)
=
-\frac{1}{2N_e}
\mathrm{Tr}
\left\{
\hat\gamma_0
[\hat u_{-\bm q},
[\hat v_{\mathrm{xc}}^{\mathrm{NL}},\hat u_{\bm q}]]
\right\}.
\label{eq:DeltaM1_trace_final}
\end{equation}

The one-particle trace in Eq.~(\ref{eq:DeltaM1_trace_final}) is
identical to the many-particle equilibrium expectation value entering
Eq.~(\ref{eq:Delta_xc}). Indeed, since
$\hat V_{\mathrm{xc}}^{\mathrm{NL}}
=\sum_i\hat v_{\mathrm{xc}}^{\mathrm{NL},(i)}$ and
$\hat\rho_{\bm q}=\sum_i\hat u_{\bm q}^{(i)}$, operators acting on
different particles commute, and hence
\begin{align}
&\left\langle
[\hat\rho_{-\bm q},
[\hat V_{\mathrm{xc}}^{\mathrm{NL}},
\hat\rho_{\bm q}]]
\right\rangle
\nonumber\\
&\qquad =
\mathrm{Tr}
\left\{
\hat\gamma_0
[\hat u_{-\bm q},
[\hat v_{\mathrm{xc}}^{\mathrm{NL}},\hat u_{\bm q}]]
\right\}.
\label{eq:one_many_body_identity}
\end{align}
Comparison with Eq.~(\ref{eq:Delta_xc}) therefore yields
\begin{equation}
\Delta M_1(\bm q)
=-\Delta_{\mathrm{xc}}(\bm q)
.
\label{eq:xc_cancellation_condition_app}
\end{equation}

Thus, local gauge covariance of the nonlocal XC potential and its
density-matrix kernel guarantees an exact cancellation between the
nonlocal XC contribution already contained in the 
noninteracting response and the corresponding kernel contribution to
the first moment. 

\section{Dynamic XC correction for the full reciprocal-space response}
\label{app:dynamic_kernel_matrix}

We generalize the dynamic correction derived in the main text to the
full reciprocal-space density-response matrix
$\chi_{\bm G\bm G'}(\bm k,\omega)$, thereby retaining local-field
effects.

For a fixed crystal momentum $\bm k$, we define
$\bm q_{\bm G}=\bm k+\bm G$. The projection introduced in
Eq.~(\ref{eq:Pq_definition}) is generalized as
\begin{equation}
\mathcal P_{\bm G\bm G'}^{\bm k}[L]
\equiv
\frac{1}{\Omega}
\int d\bm r\,d\bm x\,
e^{-i\bm q_{\bm G}\cdot\bm r}
L(\bm r,\bm r;\bm x,\bm x;\omega)
e^{i\bm q_{\bm G'}\cdot\bm x},
\label{eq:P_GG}
\end{equation}
so that
\begin{equation}
\chi_{\bm G\bm G'}(\bm k,\omega)
=
\mathcal P_{\bm G\bm G'}^{\bm k}[L](\omega).
\label{eq:chi_GG_from_L}
\end{equation}

In the asymptotic region of the upper complex-frequency plane, the
analytically continued input response has the leading form
\begin{equation}
{\boldsymbol{\chi}}_{\rm in}(\bm k,z)
=
\frac{\mathbf C^{(0)}(\bm k)}{z^2}
+\mathcal O(z^{-4}),
\label{eq:chi_matrix_hf}
\end{equation}
where
\begin{align}
C_{\bm G\bm G'}^{(0)}(\bm k)
={}&
\frac{1}{\Omega}
\sum_{\Lambda}P_\Lambda
\left\langle\Lambda\left|
[\hat\rho_{-\bm q_{\bm G}},
[\hat H_{\mathrm{KS}},
\hat\rho_{\bm q_{\bm G'}}]]
\right|\Lambda\right\rangle .
\label{eq:C0_GG}
\end{align}
For $\bm G=\bm G'$,
\begin{equation}
C_{\bm G\bm G}^{(0)}(\bm k)
=
2n_0M_1^{(0)}(\bm q_{\bm G}).
\label{eq:C0_diagonal}
\end{equation}

The density-matrix response theory changes the leading coefficient
according to
\begin{equation}
\mathbf C(\bm k)
=
\mathbf C^{(0)}(\bm k)
-
\mathbf C_{\mathrm{xc}}(\bm k),
\label{eq:C_matrix_corrected}
\end{equation}
where
\begin{align}
C_{\mathrm{xc},\bm G\bm G'}(\bm k)
={}&
\frac{1}{\Omega}
\sum_{\Lambda}P_\Lambda
\left\langle\Lambda\left|
[\hat\rho_{-\bm q_{\bm G}},
[\hat V_{\mathrm{xc}}^{\mathrm{NL}},
\hat\rho_{\bm q_{\bm G'}}]]
\right|\Lambda\right\rangle .
\label{eq:Cxc_GG}
\end{align}
In particular,
\begin{equation}
C_{\mathrm{xc},\bm G\bm G}(\bm k)
=
2n_0\Delta_{\mathrm{xc}}(\bm q_{\bm G}),
\label{eq:Cxc_diagonal}
\end{equation}
so that Eq.~(\ref{eq:C_matrix_corrected}) reduces on the diagonal to
$\Delta M_1=-\Delta_{\mathrm{xc}}$.

Making use of the general structure of the relation between an XC
kernel and the inverse density-response matrix, we write
\begin{equation}
{\boldsymbol{\chi}}^{-1}
=
{\boldsymbol{\chi}}_{\rm in}^{-1}
-
\Delta{\mathbf f}_{\mathrm{xc}} .
\label{eq:chi_corr_matrix}
\end{equation}
Since Eq.~(\ref{eq:chi_matrix_hf}) scales as $1/z^2$, the minimal
analytic correction capable of modifying its leading coefficient is
\begin{equation}
\Delta{\mathbf f}_{\mathrm{xc}}(\bm k,z)
=
z^2\mathbf F(\bm k).
\label{eq:Deltafxc_matrix_ansatz}
\end{equation}

Rearranging Eq.~(\ref{eq:chi_corr_matrix}) and using
Eqs.~(\ref{eq:chi_matrix_hf}) and
(\ref{eq:Deltafxc_matrix_ansatz}) gives, in the same asymptotic region,
\begin{equation}
{\boldsymbol{\chi}}(\bm k,z)
=
\frac{
\left[\mathbf I-\mathbf C^{(0)}(\bm k)\mathbf F(\bm k)\right]^{-1}
\mathbf C^{(0)}(\bm k)}
{z^2}
+\mathcal O(z^{-4}).
\label{eq:chi_corr_matrix_hf}
\end{equation}
Matching its leading coefficient to $\mathbf C(\bm k)$ yields
\begin{equation}
\mathbf F(\bm k)
=
\left[\mathbf C^{(0)}(\bm k)\right]^{-1}
-
\left[\mathbf C(\bm k)\right]^{-1}.
\label{eq:F_matrix}
\end{equation}
Using Eq.~(\ref{eq:C_matrix_corrected}),
\begin{equation}
\mathbf F(\bm k)
=
\left[\mathbf C^{(0)}(\bm k)\right]^{-1}
-
\left[\mathbf C^{(0)}(\bm k)
-\mathbf C_{\mathrm{xc}}(\bm k)\right]^{-1}.
\label{eq:F_matrix_Cxc}
\end{equation}

The asymptotic analysis determines the matrix coefficient
$\mathbf F(\bm k)$, while the correction at real frequencies is obtained
from the boundary value of Eq.~(\ref{eq:Deltafxc_matrix_ansatz}),
\begin{equation}
\Delta\mathbf f_{\mathrm{xc}}(\bm k,\omega)
=\omega^2\left\{\left[\mathbf C^{(0)}(\bm k)\right]^{-1}
-\left[\mathbf C^{(0)}(\bm k)-\mathbf C_{\mathrm{xc}}(\bm k)\right]^{-1}
\right\}.
\label{eq:Deltafxc_matrix_Cxc}
\end{equation}

\section{Optical-limit}
\label{sec:optical_correction}

We next consider the longitudinal optical limit $q\rightarrow0$ at
fixed nonzero frequency. Using
Eq.~(\ref{eq:Delta_ion_asymptotic}) and Eq.~(\ref{eq:dynamic_xc_correction}), for the optical-limit of the dynamic XC correction, we find
\begin{equation}
\Delta f_{\mathrm{xc}}(\bm q,\omega)
\xrightarrow[q\to0]{}
-\mathcal A_{\mathrm{xc}}\,
\frac{\omega^2}{q^2},
\label{eq:fxc_optical}
\end{equation}
with
\begin{equation}
\mathcal A_{\mathrm{xc}}
=
\frac{\widetilde{\Delta}_{\mathrm{xc}}}
{n_0
\left(1+\widetilde{\Delta}_{\mathrm{ion}}\right)
\left(1+\widetilde{\Delta}_{\mathrm{ion}}
+\widetilde{\Delta}_{\mathrm{xc}}\right)} .
\label{eq:Axc_optical}
\end{equation}

Combining Eq.~(\ref{eq:chi_corr}) with
Eq.~(\ref{eq:epsinv_chi}) and the longitudinal relation
$\epsilon(\omega)=1+4\pi i\sigma(\omega)/\omega$, and using
Eq.~(\ref{eq:fxc_optical}),  we obtain in the
optical limit
\begin{equation}
\sigma^{-1}(\omega)
=\sigma_{\rm in}^{-1}(\omega)
-i\omega\mathcal A_{\mathrm{xc}} .
\label{eq:sigma_corr_optical}
\end{equation}
Equivalently,
\begin{equation}
\sigma(\omega)
=
\frac{\sigma_{\rm in}(\omega)}
{1-i\omega\mathcal A_{\mathrm{xc}}\sigma_{\rm in}(\omega)} .
\label{eq:sigma_corr_optical_direct}
\end{equation}

If the input conductivity has a finite DC limit,
$\sigma_{\rm in}(\omega\to0)=\sigma_{\rm DC}^{\rm in}$, Eq.~(\ref{eq:sigma_corr_optical_direct})  gives $\sigma_{\rm DC}=\sigma_{\rm DC}^{\rm in}$.
Thus, the dynamic XC correction modifies the finite-frequency optical
response and its spectral weight without altering the DC limit.

The corresponding dynamic dielectric function follows directly as
\begin{equation}
\epsilon(\omega)
=1+\frac{\epsilon_{\rm in}(\omega)-1}
{1-\dfrac{\mathcal A_{\mathrm{xc}}\omega^2}{4\pi}
[\epsilon_{\rm in}(\omega)-1]} .
\label{eq:epsilon_corr_optical}
\end{equation}
Hence, although the density-space correction diverges as $1/q^2$ in the optical limit, its contribution to the dynamic conductivity and dielectric function remains finite.

The same result has a direct consequence for the
TRK sum rule for the optical conductivity.
Using Eq.~(\ref{eq:M1_exact}) together with
Eq.~(\ref{eq:Delta_ion_asymptotic}), the high-frequency expansion of
the corrected dielectric function in the optical limit reads
\begin{equation}
\epsilon(\omega)
=
1-
\frac{\omega_{\rm p,0}^{2}
\left(1+\widetilde{\Delta}_{\mathrm{ion}}\right)}
{\omega^2}
+\mathcal O(\omega^{-4}).
\label{eq:epsilon_corr_hf_optical}
\end{equation}
Consequently, using
$\epsilon(\omega)=1+4\pi i\sigma(\omega)/\omega$, the corresponding
conductivity behaves as
\begin{equation}
\sigma(\omega)
=
\frac{i n_0
\left(1+\widetilde{\Delta}_{\mathrm{ion}}\right)}
{\omega}
+\mathcal O(\omega^{-3}).
\label{eq:sigma_corr_hf_optical}
\end{equation}

On the other hand, the Kramers--Kronig relation for the optical
conductivity gives at high frequency
\begin{equation}
\sigma(\omega)
=
\frac{2i}{\pi\omega}
\int_0^\infty d\omega'\,
\operatorname{Re}\sigma(\omega')
+\mathcal O(\omega^{-3}).
\label{eq:sigma_KK_hf}
\end{equation}
Comparison of Eq.~(\ref{eq:sigma_KK_hf}) with Eq.~(\ref{eq:sigma_corr_hf_optical}) yields the TRK sum rule extensions in the case of the KSDFT simulations with non-local potentials
\begin{equation}
\int_0^\infty d\omega\,
\operatorname{Re}\sigma(\omega)
=\frac{\pi n_0}{2}
\left(1+\widetilde{\Delta}_{\mathrm{ion}}\right).
\label{eq:TRK_corr}
\end{equation}

In contrast, a conventional TDDFT calculation based on a nonlocal XC
potential, but without the derived density-matrix response correction, gives
\begin{equation}
\int_0^\infty d\omega\,
\operatorname{Re}\sigma_{\rm in}(\omega)
=\frac{\pi n_0}{2}
\left(1+\widetilde{\Delta}_{\mathrm{ion}}
+\widetilde{\Delta}_{\mathrm{xc}}
\right),
\label{eq:TRK_uncorr}
\end{equation}
where the same nonlocal-XC contribution that produces the spurious correction to the first moment of the DSF also produces a spurious
contribution to the integrated optical spectral weight. The dynamic XC correction in Eq.~(\ref{eq:fxc_optical}) removes this contribution, restoring the conductivity sum rule consistent with the density-matrix response theory.

\section{Small-wavenumber limit of the contributions due to non-local potentials to the the first moment of the DSF}\label{app:A}

To determine the long-wavelength behavior of a nonlocal contribution to the first moment, we consider a general one-electron nonlocal potential $\hat v^{\mathrm{NL}}$. We denote the corresponding contribution by $\Delta_{\mathrm{NL}}(\bm q)$; choosing $\hat v^{\mathrm{NL}}=\hat v_{\mathrm{ion}}^{\mathrm{NL}}$ gives $\Delta_{\mathrm{ion}}(\bm q)$, whereas $\hat v^{\mathrm{NL}}=\hat v_{\mathrm{xc}}^{\mathrm{NL}}$ gives $\Delta_{\mathrm{xc}}(\bm q)$.

Introducing the one-electron density translation operator
\begin{equation}
\hat n_{\bm q}^{(i)}=e^{-i\bm q\cdot\hat{\bm r}*i},
\label{eq:nq_i}
\end{equation}
the many-electron density operator can be written as
\begin{equation}
\hat\rho*{\bm q}=\sum_{i=1}^{N_e}\hat n_{\bm q}^{(i)}.
\label{eq:rho_one_electron}
\end{equation}
Likewise, the corresponding many-electron nonlocal potential is
\begin{equation}
\hat V^{\mathrm{NL}}=\sum_{i=1}^{N_e}\hat v^{\mathrm{NL},(i)}.
\label{eq:VNL_many}
\end{equation}
Since operators acting on different electrons commute, substitution of Eqs.~(\ref{eq:rho_one_electron}) and~(\ref{eq:VNL_many}) into the double commutator gives
\begin{equation}
[\hat\rho_{-\bm q},[\hat V^{\mathrm{NL}},\hat\rho_{\bm q}]]
=\sum_{i=1}^{N_e}[\hat n_{-\bm q}^{(i)},[\hat v^{\mathrm{NL},(i)},\hat n_{\bm q}^{(i)}]].
\label{eq:double_comm_reduction}
\end{equation}
The corresponding contribution to the first moment is therefore
\begin{equation}
\Delta_{\mathrm{NL}}(\bm q)=\frac{1}{2N_e}\sum_{i=1}^{N_e}\sum_{\Lambda}P_\Lambda
\left\langle\Lambda\left|[\hat n_{-\bm q}^{(i)},[\hat v^{\mathrm{NL},(i)},\hat n_{\bm q}^{(i)}]]\right|\Lambda\right\rangle .
\label{eq:Delta_NL}
\end{equation}
Eq.~(\ref{eq:Delta_NL}) shows that the $q$ dependence of $\Delta_{\mathrm{NL}}(\bm q)$ is determined by the one-electron double commutator appearing inside the many-electron matrix element.

We therefore first determine this one-electron operator explicitly. Suppressing the electron label $(i)$, a general nonlocal one-electron potential acts on a wave function according to
\begin{equation}
(\hat v^{\mathrm{NL}}\psi)(\bm r)=\int d^3r',v^{\mathrm{NL}}(\bm r,\bm r')\psi(\bm r'),
\label{eq:vnl_action}
\end{equation}
which defines its coordinate-space kernel,
\begin{equation}
\langle\bm r|\hat v^{\mathrm{NL}}|\bm r'\rangle=v^{\mathrm{NL}}(\bm r,\bm r').
\label{eq:vnl_kernel}
\end{equation}
Similarly, Eq.~(\ref{eq:nq_i}) gives
\begin{equation}
\langle\bm r|\hat n_{\bm q}|\bm r'\rangle=e^{-i\bm q\cdot\bm r}\delta(\bm r-\bm r').
\label{eq:nq_kernel}
\end{equation}

To evaluate the first commutator in Eq.~(\ref{eq:Delta_NL}), we first consider the product $\hat v^{\mathrm{NL}}\hat n_{\bm q}$. Inserting a coordinate-space resolution of the identity and using Eqs.~(\ref{eq:vnl_kernel}) and~(\ref{eq:nq_kernel}) gives
\begin{align}
\langle\bm r|\hat v^{\mathrm{NL}}\hat n_{\bm q}|\bm r'\rangle
&=\int d^3r''\,v^{\mathrm{NL}}(\bm r,\bm r'')e^{-i\bm q\cdot\bm r'}\delta(\bm r''-\bm r')
\nonumber\\
&=v^{\mathrm{NL}}(\bm r,\bm r')e^{-i\bm q\cdot\bm r'}.
\label{eq:vn_product}
\end{align}
Reversing the operator order gives
\begin{equation}
\langle\bm r|\hat n_{\bm q}\hat v^{\mathrm{NL}}|\bm r'\rangle
=e^{-i\bm q\cdot\bm r}v^{\mathrm{NL}}(\bm r,\bm r').
\label{eq:nv_product}
\end{equation}
Subtracting Eq.~(\ref{eq:nv_product}) from Eq.~(\ref{eq:vn_product}) yields
\begin{equation}
\langle\bm r|[\hat v^{\mathrm{NL}},\hat n_{\bm q}]|\bm r'\rangle
=v^{\mathrm{NL}}(\bm r,\bm r')\left(e^{-i\bm q\cdot\bm r'}-e^{-i\bm q\cdot\bm r}\right).
\label{eq:first_kernel}
\end{equation}

Using Eq.~(\ref{eq:first_kernel}) to evaluate the second commutator in Eq.~(\ref{eq:Delta_NL}) gives
\begin{equation}
\langle\bm r|[\hat n_{-\bm q},[\hat v^{\mathrm{NL}},\hat n_{\bm q}]]|\bm r'\rangle
=v^{\mathrm{NL}}(\bm r,\bm r')\left[e^{i\bm q\cdot(\bm r-\bm r')}+e^{-i\bm q\cdot(\bm r-\bm r')}-2\right].
\label{eq:double_kernel_phase}
\end{equation}
Using $e^{ix}+e^{-ix}-2=-4\sin^2(x/2)$ in Eq.~(\ref{eq:double_kernel_phase}), we obtain
\begin{equation}
\langle\bm r|[\hat n_{-\bm q},[\hat v^{\mathrm{NL}},\hat n_{\bm q}]]|\bm r'\rangle
=-4v^{\mathrm{NL}}(\bm r,\bm r')\sin^2\left[\frac{\bm q\cdot(\bm r-\bm r')}{2}\right].
\label{eq:double_kernel}
\end{equation}

Eq.~(\ref{eq:double_kernel}) gives the coordinate-space kernel of the one-electron operator entering Eq.~(\ref{eq:Delta_NL}). To connect this result back to the many-electron matrix elements in Eq.~(\ref{eq:Delta_NL}), we introduce the one-particle reduced density matrix. For a many-electron state $|\Lambda\rangle$ with wave function $\Psi_\Lambda(\bm r_1,\ldots,\bm r_{N_e})$, its one-particle density matrix is
\begin{equation}
\gamma_\Lambda(\bm r,\bm r')
=N_e\int d\vec r_2\cdots d\vec r_{N_e},
\Psi_\Lambda(\bm r,\bm r_2,\ldots)\Psi_\Lambda^*(\bm r',\bm r_2,\ldots),
\label{eq:gamma_Lambda}
\end{equation}
where spin variables are suppressed for simplicity. 

The equilibrium one-particle density matrix is the  statistical average over the states $|\Lambda\rangle$ that appears in Eq.~(\ref{eq:Delta_NL}),
\begin{equation}
\gamma(\bm r,\bm r')=\sum_\Lambda P_\Lambda,\gamma_\Lambda(\bm r,\bm r').
\label{eq:gamma_eq}
\end{equation}

For any one-electron operator $\hat O$ with kernel $O(\bm r,\bm r')=\langle\bm r|\hat O|\bm r'\rangle$, Eqs.~(\ref{eq:gamma_Lambda}) and~(\ref{eq:gamma_eq}) imply
\begin{equation}
\sum_{\Lambda}P_\Lambda\sum_{i=1}^{N_e}\langle\Lambda|\hat O^{(i)}|\Lambda\rangle
=\int d\vec r d\vec r'\,\gamma(\bm r',\bm r)O(\bm r,\bm r').
\label{eq:onebody_expectation_gamma}
\end{equation}

Applying Eq.~(\ref{eq:onebody_expectation_gamma}) to the double-commutator kernel in Eq.~(\ref{eq:double_kernel}) converts Eq.~(\ref{eq:Delta_NL}) into
\begin{align}
\Delta_{\mathrm{NL}}(\bm q)
&=-\frac{2}{N_e}\int d\vec r\,d\vec r'\,
\gamma(\bm r',\bm r)v^{\mathrm{NL}}(\bm r,\bm r')
\nonumber\\
&\qquad\times
\sin^2\left[\frac{\bm q\cdot(\bm r-\bm r')}{2}\right].
\label{eq:Delta_NL_coordinate}
\end{align}
Eq.~(\ref{eq:Delta_NL_coordinate}) is the coordinate-space form of the original definition in Eq.~(\ref{eq:Delta_NL}).

Writing $\bm q=q\widehat{\bm e}_{\bm q}$, the pointwise small-$q$ limit of the factor in Eq.~(\ref{eq:Delta_NL_coordinate}) is
\begin{equation}
\lim_{q\rightarrow0}\frac{4\sin^2[q\widehat{\bm e}_{\bm q}\cdot(\bm r-\bm r')/2]}{q^2}
=[\widehat{\bm e}_{\bm q}\cdot(\bm r-\bm r')]^2.
\label{eq:sin_q_limit}
\end{equation}

Eq.~(\ref{eq:sin_q_limit}) alone does not guarantee that the limit can be taken inside the coordinate integrals in Eq.~(\ref{eq:Delta_NL_coordinate}). A sufficient condition follows from the inequality $|\sin x|\leq|x|$, which gives
\begin{equation}
\frac{4\sin^2[q\widehat{\bm e}_{\bm q}\cdot(\bm r-\bm r')/2]}{q^2}
\leq[\widehat{\bm e}_{\bm q}\cdot(\bm r-\bm r')]^2\leq|\bm r-\bm r'|^2.
\label{eq:sin_bound}
\end{equation}
According to Eq.~(\ref{eq:sin_bound}), the integrand of $\Delta_{\mathrm{NL}}(q\widehat{\bm e}_{\bm q})/q^2$ is bounded by an integrable $q$-independent function if
\begin{equation}
\int  d\vec r d\vec r'\,
\left|\gamma(\bm r',\bm r)v^{\mathrm{NL}}(\bm r,\bm r')\right|
|\bm r-\bm r'|^2<\infty .
\label{eq:nonlocal_second_moment}
\end{equation}
Eq.~(\ref{eq:nonlocal_second_moment}) is therefore a sufficient condition for the dominated-convergence theorem to be applied to Eq.~(\ref{eq:Delta_NL_coordinate}). Physically, it requires the nonlocal potential, weighted by the equilibrium one-particle density matrix, to have a finite second moment with respect to the nonlocal separation $|\bm r-\bm r'|$.

Under the condition in Eq.~(\ref{eq:nonlocal_second_moment}),
Eqs.~(\ref{eq:Delta_NL_coordinate}) and~(\ref{eq:sin_q_limit}) show
that the ratio $\Delta_{\mathrm{NL}}(\vec q)/q^2$ approaches a
finite limit. We denote this long-wavelength coefficient by
\begin{equation}
\widetilde{\Delta}_{\mathrm{NL}}(\widehat{\bm e}_{\bm q})
\equiv
\lim_{q\rightarrow0}
\frac{\Delta_{\mathrm{NL}}(q\widehat{\bm e}_{\bm q})}{q^2}.
\label{eq:Delta_NL_tilde_def}
\end{equation}
Using Eq.~(\ref{eq:Delta_NL_coordinate}), $\widetilde{\Delta}_{\mathrm{NL}}(\widehat{\bm e}_{\bm q})$ is given explicitly by
\begin{align}
\widetilde{\Delta}_{\mathrm{NL}}(\widehat{\bm e}_{\bm q})
&=-\frac{1}{2N_e}
\int d\vec r\,d\vec r'\,
\gamma_{\mathrm{eq}}(\bm r',\bm r)
v^{\mathrm{NL}}(\bm r,\bm r')
\nonumber\\
&\qquad\times
[\widehat{\bm e}_{\bm q}\cdot(\bm r-\bm r')]^2 .
\label{eq:Delta_NL_tilde}
\end{align}
The finiteness of Eq.~(\ref{eq:Delta_NL_tilde}) then implies the
long-wavelength expansion
\begin{equation}
\Delta_{\mathrm{NL}}(q\widehat{\bm e}_{\bm q})
=
q^2\widetilde{\Delta}_{\mathrm{NL}}(\widehat{\bm e}_{\bm q})
+o(q^2),
\qquad q\rightarrow0.
\label{eq:Delta_NL_asymptotic}
\end{equation}

Thus, the quadratic long-wavelength behavior in Eq.~(\ref{eq:Delta_NL_asymptotic}) is not specific to a particular pseudopotential or XC potential. It follows for a general one-electron nonlocal potential provided that the finite-second-moment condition in Eq.~(\ref{eq:nonlocal_second_moment}) is satisfied. In particular, this result applies to $\Delta_{\mathrm{ion}}(\bm q)$ and $\Delta_{\mathrm{xc}}(\bm q)$ when the corresponding nonlocal electron--ion and XC kernels satisfy this condition.

\section{Dynamic electron--ion correction in OF-TDDFT}
\label{app:OF_TDDFT_dynamic_ei}

In orbital-free DFT (OFDFT), a standard strategy is to use known
properties of the linear density response to construct approximations
for the required potentials, both in equilibrium
\cite{Michele_Chem_Rev,Moldabekov_PRB_2023,Moldabekov_Electronic_Structure, SKANEX}
and in time-dependent formulations
\cite{White_PRB_2018,Jiang_PRB,Moldabekov_pop_2018}.
Following the same strategy, we first derive the correction required
at the level of the dynamic density response and then construct the
corresponding effective electron--ion potential.

For a local pseudopotential,
$\Delta_{\rm ion}(\bm q)=0$. To restore the contribution associated
with a nonlocal electron--ion interaction to the first moment, we
introduce the dynamic kernel
\begin{equation}
\Delta f_{\mathrm{ei}}(\bm q,z)
=
\chi_{\rm lpp}^{-1}(\bm q,z)
-
\chi_{\rm nlpp}^{-1}(\bm q,z),
\label{eq:chi_corr_ofdft}
\end{equation}
where $\chi_{\rm lpp}(\bm q,z)$ denotes the density response obtained
with a local pseudopotential, whereas
$\chi_{\rm nlpp}(\bm q,z)$ denotes the response corresponding to a
nonlocal pseudopotential for which
$\Delta_{\rm ion}(\bm q)\neq0$.

Based on  Eq.~(\ref{eq:chi_corr_ofdft}), following steps analogous to those used in
Sec.~\ref{subsec:non_imp_xc} to derive the dynamic correction to the
XC kernel and using the long-wavelength limit of  $\Delta_{\rm ion}(\bm q)$, Eq.~(\ref{eq:Delta_ion_asymptotic}), we obtain
\begin{equation}
\Delta f_{\mathrm{ei}}(\bm q,\omega)
=
\frac{\widetilde{\Delta}_{\mathrm{ei}}}
{n_0\left[1/2+\widetilde{\Delta}_{\mathrm{ei}}\right]}
\frac{\omega^2}{q^2}.
\label{eq:app_dynamic_ion_kernel}
\end{equation}
For compactness, we define
\begin{equation}
C_{\mathrm{ei}}
=
\frac{\widetilde{\Delta}_{\mathrm{ei}}}
{n_0\left[1/2+\widetilde{\Delta}_{\mathrm{ei}}\right]},
\label{eq:app_Cei}
\end{equation}
so that
\begin{equation}
\Delta f_{\mathrm{ei}}(\bm q,\omega)
=
C_{\mathrm{ei}}\frac{\omega^2}{q^2}.
\label{eq:app_kernel_compact}
\end{equation}

The corresponding effective nonlocal electron--ion potential is
defined in linear response through
\begin{equation}
\delta V_{\mathrm{ei}}^{\rm NL}(\bm q,\omega)
=
\Delta f_{\mathrm{ei}}(\bm q,\omega)
\delta n(\bm q,\omega),
\label{eq:app_kernel_definition}
\end{equation}
and therefore
\begin{equation}
\delta V_{\mathrm{ei}}^{\rm NL}(\bm q,\omega)
=
C_{\mathrm{ei}}
\frac{\omega^2}{q^2}
\delta n(\bm q,\omega).
\label{eq:app_Vei_freq}
\end{equation}

Using
$\partial_t^2\leftrightarrow-\omega^2$ and
$\nabla^2\leftrightarrow-q^2$, the corresponding real-space
representation can be written in the particularly simple form
\begin{equation}
\nabla^2 V_{\mathrm{ei}}^{\rm NL}(\bm r,t)
=
C_{\mathrm{ei}}
\frac{\partial^2 n(\bm r,t)}{\partial t^2}.
\label{eq:app_Vei_realspace}
\end{equation}
Thus, $V_{\mathrm{ei}}^{\rm NL}(\bm r,t)$ is obtained from a
Poisson-type equation whose source is proportional to the second time
derivative of the electronic density. Fourier transformation of
Eq.~(\ref{eq:app_Vei_realspace}) gives
\begin{equation}
-q^2\,
\delta V_{\mathrm{ei}}^{\rm NL}(\bm q,\omega)
=
-C_{\mathrm{ei}}\omega^2
\delta n(\bm q,\omega),
\end{equation}
which immediately recovers Eq.~(\ref{eq:app_Vei_freq}).

For a periodic system and a perturbation characterized by the Bloch
wave vector $\bm q$, the density variation can be expanded as
\begin{equation}
\delta n(\bm r,t)
=
\sum_{\bm G}
\delta n_{\bm G}(\bm q,t)
e^{i(\bm q+\bm G)\cdot\bm r},
\label{eq:app_density_G}
\end{equation}
where $\bm G$ denotes a reciprocal-lattice vector. Since
\begin{equation}
\nabla^2
e^{i(\bm q+\bm G)\cdot\bm r}
=
-|\bm q+\bm G|^2
e^{i(\bm q+\bm G)\cdot\bm r},
\end{equation}
Eq.~(\ref{eq:app_Vei_realspace}) yields
\begin{equation}
V_{\mathrm{ei},\bm G}^{\rm NL}(\bm q,t)
=
-C_{\mathrm{ei}}
\frac{1}{|\bm q+\bm G|^2}
\frac{\partial^2
\delta n_{\bm G}(\bm q,t)}
{\partial t^2}.
\label{eq:app_Vei_G}
\end{equation}
The spatially nonlocal operation therefore reduces to a diagonal
multiplication by $|\bm q+\bm G|^{-2}$ in reciprocal space.

For the macroscopic component, $\bm G=0$,
Eq.~(\ref{eq:app_Vei_G}) becomes
\begin{equation}
V_{\mathrm{ei},0}^{\rm NL}(\bm q,t)
=
-C_{\mathrm{ei}}
\frac{1}{q^2}
\frac{\partial^2
\delta n_0(\bm q,t)}
{\partial t^2}.
\label{eq:app_Vei_head}
\end{equation}
Fourier transformation with respect to time then gives
\begin{equation}
\delta V_{\mathrm{ei},0}^{\rm NL}(\bm q,\omega)
=
C_{\mathrm{ei}}
\frac{\omega^2}{q^2}
\delta n_0(\bm q,\omega),
\end{equation}
which is precisely the correction generated by
Eq.~(\ref{eq:app_dynamic_ion_kernel}).

In real-time OF-TDDFT, the electronic density is commonly represented
in terms of a single effective orbital,
\begin{equation}
\Phi(\bm r,t)
=
\sqrt{n(\bm r,t)}
e^{iS(\bm r,t)},
\label{eq:app_OF_orbital}
\end{equation}
whose propagation can be written schematically as \cite{Jiang_PRB_2021, SciPostPhys.12.2.062}
\begin{equation}
i\frac{\partial\Phi(\bm r,t)}{\partial t}
=
\left[
-\frac{\nabla^2}{2}
+
V_{\mathrm{OF}}[n](\bm r,t)
\right]
\Phi(\bm r,t),
\label{eq:app_OF_standard}
\end{equation}
where atomic units are used and $V_{\mathrm{OF}}[n]$ contains the
external, Hartree, local electron--ion, XC, and orbital-free kinetic
contributions appropriate to the particular OF-TDDFT formulation.
The missing effect associated with the nonlocal electron--ion
interaction can then be incorporated through
\begin{equation}
i\frac{\partial\Phi(\bm r,t)}{\partial t}
=
\left[
-\frac{\nabla^2}{2}
+
V_{\mathrm{OF}}[n](\bm r,t)
+
V_{\mathrm{ei}}^{\rm NL}[n](\bm r,t)
\right]
\Phi(\bm r,t),
\label{eq:app_OF_corrected}
\end{equation}
where $V_{\mathrm{ei}}^{\rm NL}[n](\bm r,t)$ is determined from
Eq.~(\ref{eq:app_Vei_realspace}).

For a plane-wave implementation of real-time OF-TDDFT,
Eq.~(\ref{eq:app_Vei_G}) provides a particularly convenient
representation. At each time step, the density is transformed to
reciprocal space, its second time derivative is evaluated, and each
Fourier component is multiplied by
$-C_{\mathrm{ei}}/|\bm q+\bm G|^2$. The resulting
$V_{\mathrm{ei}}^{\rm NL}$ is then transformed back to real space
before the subsequent propagation step. Hence, despite representing
a spatially nonlocal electron--ion contribution, its numerical
evaluation requires an operation closely analogous to the
reciprocal-space solution of the Poisson equation routinely used for
the Hartree potential.

%on the HoreKa supercomputer funded by the Ministry of
%Science, Research and the Arts Baden-W\"urttemberg and by the Federal
%Ministry of Education and Research.

% ============================================================
% Author contributions
% ============================================================

%\section*{Author Contributions}

% Example:
% Z.A.M. conceived the study and performed ...
% M.P. ...
% J.V. ...
% T.D. ...
% All authors discussed the results and contributed to the manuscript.

% ============================================================
% Competing interests
% ============================================================

\section*{Competing Interests}

The authors declare no competing interests.

% ============================================================
% References
% ============================================================
\bibliography{bibliography}
\end{document}